# On the Electrodynamics of Moving Permanent Dipoles in External Electromagnetic Fields

Masud Mansuripur
College of Optical Sciences, The University of Arizona, Tucson

**Abstract**. The classical theory of electrodynamics is built upon Maxwell's equations and the concepts of electromagnetic field, force, energy and momentum, which are intimately tied together by Poynting's theorem and the Lorentz force law. Whereas Maxwell's macroscopic equations relate the electric and magnetic fields to their material sources (i.e., charge, current, polarization and magnetization), Poynting's theorem governs the flow of electromagnetic energy and its exchange between fields and material media, while the Lorentz law regulates the back-and-forth transfer of momentum between the media and the fields. The close association of momentum with energy thus demands that the Poynting theorem and the Lorentz law remain consistent with each other, while, at the same time, ensuring compliance with the conservation laws of energy, linear momentum, and angular momentum. This paper shows how a consistent application of the aforementioned laws of electrodynamics to moving permanent dipoles (both electric and magnetic) brings into play the rest-mass of the dipoles. The rest mass must vary in response to external electromagnetic fields if the overall energy of the system is to be conserved. The physical basis for the inferred variations of the rest-mass appears to be an interference between the internal fields of the dipoles and the externally applied fields. We use two different formulations of the classical theory in which energy and momentum relate differently to the fields, yet we find identical behavior for the rest-mass in both formulations.

**1. Introduction**. The electrodynamics of moving media is a complex subject that has been discussed in several papers, textbooks and monographs [1-18], yet continues to attract attention for its practical applications as well as its relevance to fundamental issues involving field-matter interactions. In the early days of the 20$^{th}$ century, Minkowski used ideas from the newly developed theory of relativity to analyze the dynamics of arbitrarily moving bodies in the presence of electromagnetic (EM) fields. He constructed a stress-energy tensor for EM systems that incorporated the electric field $E$, the magnetic field $H$, the displacement $D$, and the magnetic induction $B$, relating the strength of these fields at each point $(r, t)$ in space-time to the corresponding field strengths in a local rest frame of the material media at $(r', t')$ [1]. Einstein and Laub initially endorsed Minkowski's analysis, and proceeded to summarize his results while presenting them in a less abstract language and in simplified form [2]. Subsequently, Einstein and Laub expressed skepticism of Minkowski's general formula for the ponderomotive force exerted on bodies in the EM field, and presented their own formulation for bodies at rest [3]. In the meantime, Abraham, also relying on notions of relativity and the Lorentz transformation of coordinates and fields between inertial frames, developed a modified version of Minkowski's stress-energy tensor for moving bodies in the presence of EM fields [4,5]. In the 1950s, L. J. Chu and his colleagues at the Massachusetts Institute of Technology attempted to construct a consistent theory of macroscopic electrodynamics, and proceeded to extend their methodology to systems of moving bodies hosting electrical charges and currents as well as polarization and magnetization [6]. A good example of the sophisticated tools and techniques that have been deployed to study the electrodynamics of moving media may be found in the monograph by Penfield and Haus, who employed the principle of virtual power to derive energy, momentum, force and torque formulas for arbitrarily moving polarizable and magnetizable media in diverse circumstances [8]. Despite the progress made, Schieber's remarks on the difficulty of extending Maxwell's electrodynamics of media at rest to moving bodies are particularly instructive [12].

Compared to what has already been discussed in the literature, the present paper has a decidedly narrow focus and limited scope. However, our main conclusion that *the rest-mass of electric and magnetic dipoles must vary in the presence of external electromagnetic fields* is apparently new, as it did not turn up in our search of the existing literature. To guide the reader along the path to the above conclusion, we examine several thought experiments involving the exchange of energy and momentum between EM fields and point-particles in uniform translational or rotational motion. We rely solely on Maxwell's equations, the special theory of relativity, and the force and torque laws of classical electrodynamics. No significant approximations will be made and no new physical principles will be introduced. The reader may disagree with our interpretation of the various terms appearing in the classical equations, but the final results are independent of such interpretations and should stand on their own. We conclude that, while a point-charge has a fixed rest-mass which does not change as a result of interaction with external EM fields, permanent point-dipoles (both electric and magnetic) exhibit variations in their rest-mass which depend on their state of motion, their orientation relative to external fields, and the strength of the external fields.

After a brief overview in Sec. 2, we describe in Secs. 3 through 6 the fundamental assumptions and equations of classical electrodynamics on which the remainder of the paper is based. We take Maxwell's macroscopic equations seriously, and rely on their prescription for the EM structure of individual point-dipoles. In a departure from the conventional treatment of polarization $\boldsymbol{P}(\boldsymbol{r},t)$ and magnetization $\boldsymbol{M}(\boldsymbol{r},t)$ as locally averaged densities of dipole moments which are reducible to electrical charge and current densities, we consider $\boldsymbol{P}$ and $\boldsymbol{M}$ as precise descriptors of electric and magnetic dipole-moment densities, on a par with free charge and free current densities, $\rho_{\text{free}}(\boldsymbol{r},t)$ and $\boldsymbol{J}_{\text{free}}(\boldsymbol{r},t)$. Maxwell's macroscopic equations are thus taken to be exact mathematical relations between the primary sources ($\rho_{\text{free}}, \boldsymbol{J}_{\text{free}}, \boldsymbol{P}, \boldsymbol{M}$) and the EM fields $\boldsymbol{E}, \boldsymbol{H}, \boldsymbol{D} = \varepsilon_0 \boldsymbol{E} + \boldsymbol{P}$, and $\boldsymbol{B} = \mu_0 \boldsymbol{H} + \boldsymbol{M}$, which are produced by these sources. Maxwell's equations can then be "arranged" in different ways; we explore two such arrangements that are frequently encountered in the literature, each with its own expressions for the EM energy-density, Poynting vector, stress tensor, force and torque densities (exerted by the fields on material media), and the EM momentum-density. These two approaches to classical electrodynamics will be referred to as the Lorentz and Einstein-Laub formulations. An important conclusion reached in this paper is that, in every situation examined and despite numerous differences in the intermediate steps, the two formulations yield precisely the same answer for the time-rate-of-change of a dipole's rest-mass.

Section 7 provides a brief review of the relativistic dynamics of a point-mass. A point-charge traveling at an essentially constant velocity in a uniform electric field is examined in Sec. 8, where it is shown that the particle's rest-mass remains unchanged as it gains or loses kinetic energy. The same is *not* necessarily true of electric and magnetic point-dipoles traveling in EM fields, as shown by several examples in Secs. 9 through 12. The cases of rotating electric and magnetic dipoles in static fields are treated in Secs. 13 and 14, respectively, where it is shown once again that a dipole's rest mass (as well as its moment of inertia) could vary in the presence of an external field. The final section is devoted to closing remarks and general conclusions.

It must be emphasized at the outset that we are *not* dealing in this paper with elementary particles which obey the laws of quantum electrodynamics. Our point-particles are small spherical objects subject to Maxwell's equations and other rules and constraints of classical electrodynamics. For mathematical convenience, we rely on Dirac's delta-function to represent the localization of point-particles in three-dimensional Euclidean space. The end results, however, could be derived for any sufficiently small and homogeneous spherical particle within which the externally applied EM fields would remain substantially uniform.

**2. Synopsis**. For a free particle of rest-mass $\mathfrak{m}_0$ moving with velocity $\boldsymbol{V}(t)$, the relativistic energy and linear momentum are given, respectively, by $\mathcal{E}(t) = \gamma \mathfrak{m}_0 c^2$ and $\boldsymbol{\mathcal{P}}(t) = \gamma \mathfrak{m}_0 \boldsymbol{V}$, where $t$ stands for time, $c$ is the speed of light in vacuum, and $\gamma = 1/\sqrt{1-(V/c)^2}$. The classical EM theory enables us to calculate $\mathrm{d}\mathcal{E}/\mathrm{d}t$ and $\mathrm{d}\boldsymbol{\mathcal{P}}/\mathrm{d}t$ from a knowledge of the external fields and the EM properties of the moving particle. The close connection between the above entities then allows one to infer the time-rate-of-change of the particle's rest-mass in various circumstances.

Intuitively, one may imagine a point-charge as a small, uniformly charged, solid sphere, whose internal *E*-field points radially toward or away from the center of the particle. Imposition of an external *E*-field over this internal *E*-field does *not* affect the particle's internal energy, as the volume integral of the *E*-field energy-density, $\tfrac{1}{2}\varepsilon_0|\boldsymbol{E}_\text{ext} + \boldsymbol{E}_\text{int}|^2$, will have no contributions from the cross-term $\varepsilon_0 \boldsymbol{E}_\text{ext} \cdot \boldsymbol{E}_\text{int}$ (here $\varepsilon_0$ is the permittivity of free space). In contrast, a moving electric or magnetic dipole will have strong internal *E* and/or *H* fields which do *not* average out to zero. Superposition of internal and external fields can thus produce cross-terms in the form of $\varepsilon_0 \boldsymbol{E}_\text{ext} \cdot \boldsymbol{E}_\text{int}$ or $\mu_0 \boldsymbol{H}_\text{ext} \cdot \boldsymbol{H}_\text{int}$ or $\mu_0^{-1} \boldsymbol{B}_\text{ext} \cdot \boldsymbol{B}_\text{int}$ (with $\mu_0$ being the permeability of free space), which contribute to the particle's internal energy and, consequently, modify its rest-mass $\mathfrak{m}_0$. We present several examples in Secs. 9-14 where such situations arise, and proceed to relate the time-rate-of-change of $\mathfrak{m}_0$ to the various parameters of the system under consideration.

Our basic assumption will be that mass and energy are equivalent ($\mathcal{E} = \mathfrak{m}c^2$). A rocket or a planet shedding or accreting mass must have a time-dependent rest-mass, and so could a particle that is radiating or absorbing electromagnetic energy. Drawing a closed surface around and immediately outside the particle, we proceed to calculate the rates at which EM energy and EM momentum cross this surface and, therefore, enter or exit the particle. Once inside the particle, we suppose that the time-rate-of-change of the EM momentum, $\mathrm{d}\boldsymbol{\mathcal{P}}/\mathrm{d}t$, is the total instantaneous force $\boldsymbol{F}_\text{ext}(t)$ exerted by the external world on the moving particle. (This is the most general definition of force, although, as a matter of fact, the symbol $\boldsymbol{F}_\text{ext}$ is redundant; the entire calculation can be carried out in terms of $\mathrm{d}\boldsymbol{\mathcal{P}}/\mathrm{d}t$, without any reference whatsoever to force.) A similar argument applies to the EM energy: once it crosses the enclosing surface, it becomes part and parcel of the mass-energy of the particle. Having determined $\mathrm{d}\boldsymbol{\mathcal{P}}/\mathrm{d}t$ and $\mathrm{d}\mathcal{E}/\mathrm{d}t$ as the rates of exchange of EM momentum and EM energy between the particle and the outside world, we proceed to calculate the time-rate-of-change $\mathrm{d}(\mathfrak{m}_0 c^2)/\mathrm{d}t$ of the rest-energy of the particle using the close connection among the aforementioned entities. The exact relationship needed for this analysis will be derived in Sec. 7.

Whereas Maxwell's equations are unique and undisputed, there exist alternative expressions for EM force, torque, energy, and momentum in the classical literature. We focus our attention on two different approaches to the latter aspects of electrodynamics, one that can loosely be associated with the name of H. A. Lorentz, and another whose origins could be traced to A. Einstein and J. Laub. While in the Lorentz approach electric and magnetic dipoles are reduced to bound electrical charges and currents, the Einstein-Laub treatment considers dipoles as independent entities, on a par with free electrical charge and current. It will be seen that the Lorentz and Einstein-Laub formalisms, although differing in intermediate steps, always end up with the same results for the time-rate-of-change of the rest-mass. We have discussed the relative merits of these two formulations elsewhere [19-24], and been criticized for not paying due attention to hidden momentum in our treatment of the Lorentz formalism [25-36]. Here we only would like to emphasize that, in applications involving solid (as opposed to deformable) media, the Einstein-Laub theory yields results that are identical to those obtained in the Lorentz approach, albeit *without* the need for hidden energy and hidden momentum inside magnetic matter. In contrast, hidden entities are inescapable companions of the Lorentz approach [37-59].

**3. Maxwell's macroscopic equations**. We take Maxwell's macroscopic equations as our point of departure. In addition to free charge and free current densities, $\rho_{\text{free}}(r,t)$ and $J_{\text{free}}(r,t)$, the macroscopic equations contain polarization $P(r,t)$ and magnetization $M(r,t)$ as sources of the EM field [60-62]. It is important to recognize that Maxwell's equations, taken at face value, do not make any assumptions about the nature of $P$ and $M$, nor about the constitutions of electric and magnetic dipoles. These equations simply take polarization and magnetization as they exist in Nature, and enable one to calculate the fields $E(r,t)$ and $H(r,t)$ whenever and wherever the spatio-temporal distributions of the sources ($\rho_{\text{free}}, J_{\text{free}}, P, M$) are fully specified.

In Maxwell's macroscopic equations, $P$ and $M$ are lumped together with the $E$ and $H$ fields, which then appear as the displacement $D(r,t) = \varepsilon_o E + P$ and the magnetic induction $B(r,t) = \mu_o H + M$. In their most general form, Maxwell's macroscopic equations are written

$$\nabla \cdot D = \rho_{\text{free}}, \tag{1a}$$

$$\nabla \times H = J_{\text{free}} + \partial D/\partial t, \tag{1b}$$

$$\nabla \times E = -\partial B/\partial t, \tag{1c}$$

$$\nabla \cdot B = 0. \tag{1d}$$

It is possible to interpret the above equations in different ways, without changing the results of calculations. In what follows, we rely on two different interpretations of the macroscopic equations. This is done by simply re-arranging the equations without changing their physical content. We shall refer to the two re-arrangements (and corresponding interpretations) as the Lorentz formalism and the Einstein-Laub formalism.

In the Lorentz formalism, Eqs.(1a) and (1b) are re-organized by eliminating the $D$ and $H$ fields. The re-arranged equations are subsequently written as follows:

$$\varepsilon_o \nabla \cdot E = \rho_{\text{free}} - \nabla \cdot P, \tag{2a}$$

$$\nabla \times B = \mu_o (J_{\text{free}} + \frac{\partial P}{\partial t} + \mu_o^{-1} \nabla \times M) + \mu_o \varepsilon_o \frac{\partial E}{\partial t}, \tag{2b}$$

$$\nabla \times E = -\frac{\partial B}{\partial t}, \tag{2c}$$

$$\nabla \cdot B = 0. \tag{2d}$$

In this interpretation, electric dipoles appear as bound electric charge and current densities ($-\nabla \cdot P$ and $\partial P/\partial t$), while magnetic dipoles appear to act as Amperian current loops with a bound current-density given by $\mu_o^{-1} \nabla \times M$. None of this says anything about the physical nature of the dipoles, and whether, in reality, electric dipoles are a pair of positive and negative electric charges joined by a short spring, or whether magnetic dipoles are small, stable loops of electrical current. All one can say is that eliminating $\boldsymbol{D}$ and $\boldsymbol{H}$ from Maxwell's equations has led to a particular form of these equations which is consistent with the above suppositions concerning the physical nature of the dipoles.

Next, consider an alternative arrangement of Maxwell's equations, one that may be designated as the departure point for the Einstein-Laub formulation. Eliminating $\boldsymbol{D}$ and $\boldsymbol{B}$ from Eqs.(1), we arrive at

$$\varepsilon_0 \boldsymbol{\nabla} \cdot \boldsymbol{E} = \rho_{\text{free}} - \boldsymbol{\nabla} \cdot \boldsymbol{P}, \tag{3a}$$

$$\boldsymbol{\nabla} \times \boldsymbol{H} = \left(\boldsymbol{J}_{\text{free}} + \frac{\partial \boldsymbol{P}}{\partial t}\right) + \varepsilon_0 \frac{\partial \boldsymbol{E}}{\partial t}, \tag{3b}$$

$$\boldsymbol{\nabla} \times \boldsymbol{E} = -\frac{\partial \boldsymbol{M}}{\partial t} - \mu_0 \frac{\partial \boldsymbol{H}}{\partial t}, \tag{3c}$$

$$\mu_0 \boldsymbol{\nabla} \cdot \boldsymbol{H} = -\boldsymbol{\nabla} \cdot \boldsymbol{M}. \tag{3d}$$

In the Einstein-Laub interpretation, the electric dipoles appear as a pair of positive and negative electric charges tied together by a short spring (exactly as in the Lorentz formalism). However, each magnetic dipole now behaves as a pair of north and south poles joined by a short spring. In other words, magnetism is no longer associated with an electric current density, but rather with bound magnetic charge and magnetic current densities $-\boldsymbol{\nabla} \cdot \boldsymbol{M}$ and $\partial \boldsymbol{M}/\partial t$. We emphasize once again that such interpretations have nothing to do with the physical reality of the dipoles. The north and south poles mentioned above are not necessarily magnetic monopoles (i.e., the so-called Gilbert model [29]); rather, they are "fictitious" charges that acquire meaning only when Maxwell's equations are written in the form of Eqs.(3).

The two forms of Maxwell's equations given by Eqs.(2) and (3) are identical, in the sense that, given the source distributions $(\rho_{\text{free}}, \boldsymbol{J}_{\text{free}}, \boldsymbol{P}, \boldsymbol{M})$, these two sets of equations predict exactly the same EM fields $(\boldsymbol{E}, \boldsymbol{D}, \boldsymbol{B}, \boldsymbol{H})$ throughout space and time. How one chooses to "interpret" the physical nature of the dipoles is simply a matter of taste and personal preference. Such interpretations are totally irrelevant as far as the solutions of Maxwell's equations are concerned.

**4. Electromagnetic energy**. Different interpretations of Maxwell's macroscopic equations lead to different expressions for the EM energy-density and energy flow-rate. However, as will be seen in the examples of the following sections, the end results turn out to be the same.

In the Lorentz formulation, we dot-multiply $\boldsymbol{B}(\boldsymbol{r}, t)$ into Eq.(2c), then subtract it from the dot-product of $\boldsymbol{E}(\boldsymbol{r}, t)$ and Eq.(2b). The end result is

$$\boldsymbol{E} \cdot \boldsymbol{\nabla} \times \boldsymbol{B} - \boldsymbol{B} \cdot \boldsymbol{\nabla} \times \boldsymbol{E} = \mu_0 \boldsymbol{E} \cdot \left(\boldsymbol{J}_{\text{free}} + \frac{\partial \boldsymbol{P}}{\partial t} + \mu_0^{-1} \boldsymbol{\nabla} \times \boldsymbol{M}\right) + \mu_0 \varepsilon_0 \boldsymbol{E} \cdot \frac{\partial \boldsymbol{E}}{\partial t} + \boldsymbol{B} \cdot \frac{\partial \boldsymbol{B}}{\partial t}. \tag{4}$$

The left-hand-side of the above equation is equal to $-\boldsymbol{\nabla} \cdot (\boldsymbol{E} \times \boldsymbol{B})$ according to a well-known vector identity. We may thus *define* the Poynting vector in the Lorentz formalism as

$$\boldsymbol{S}_L = \mu_0^{-1} \boldsymbol{E} \times \boldsymbol{B}, \tag{5}$$

and proceed to rewrite Eq.(4) as

$$\boldsymbol{\nabla} \cdot \boldsymbol{S}_L + \frac{\partial}{\partial t}(\tfrac{1}{2}\varepsilon_0 \boldsymbol{E} \cdot \boldsymbol{E} + \tfrac{1}{2}\mu_0^{-1} \boldsymbol{B} \cdot \boldsymbol{B}) + \boldsymbol{E} \cdot \left(\boldsymbol{J}_{\text{free}} + \frac{\partial \boldsymbol{P}}{\partial t} + \mu_0^{-1} \boldsymbol{\nabla} \times \boldsymbol{M}\right) = 0. \tag{6}$$

Thus, in the Lorentz interpretation, EM energy flows at a rate of $S_L$ per unit area per unit time, the stored energy-density in the $E$ and $B$ fields is

$$\mathcal{E}_L(r,t) = \tfrac{1}{2}\varepsilon_0 E \cdot E + \tfrac{1}{2}\mu_0^{-1} B \cdot B, \tag{7}$$

and energy is exchanged between fields and media at a rate of

$$\tfrac{\partial}{\partial t}\mathcal{E}_L^{(\text{exch})}(r,t) = E \cdot J_{\text{total}} \quad \text{(per unit volume per unit time)}, \tag{8}$$

where $J_{\text{total}}$ is the sum of free and bound current densities; see Eq.(6). Note that the exchange of energy between the fields and the media is a two-way street: When $E \cdot J$ is positive, energy leaves the field and enters the material medium, and when $E \cdot J$ is negative, energy flows in the opposite direction. All in all, we have imposed our own interpretation on the various terms appearing in Eq.(6), which is the mathematical expression of energy conservation. The validity of Eq.(6), however, being a direct and rigorous consequence of Maxwell's equations, is independent of any specific interpretation.

A similar treatment of EM energy-density and energy flow-rate can be carried out in the Einstein-Laub approach. This time, we dot-multiply Eq.(3c) into $H(r,t)$ and subtract the resulting equation from the dot-product of $E(r,t)$ into Eq.(3b). We find

$$\nabla \cdot (E \times H) + \tfrac{\partial}{\partial t}(\tfrac{1}{2}\varepsilon_0 E \cdot E + \tfrac{1}{2}\mu_0 H \cdot H) + \left(E \cdot J_{\text{free}} + E \cdot \tfrac{\partial P}{\partial t} + H \cdot \tfrac{\partial M}{\partial t}\right) = 0. \tag{9}$$

Thus, in the Einstein-Laub interpretation, the Poynting vector is

$$S_{EL} = E \times H, \tag{10}$$

the stored energy-density in the $E$ and $H$ fields is

$$\mathcal{E}_{EL}(r,t) = \tfrac{1}{2}\varepsilon_0 E \cdot E + \tfrac{1}{2}\mu_0 H \cdot H, \tag{11}$$

and energy is exchanged between fields and media at the rate of

$$\tfrac{\partial}{\partial t}\mathcal{E}_{EL}^{(\text{exch})}(r,t) = E \cdot J_{\text{free}} + E \cdot \tfrac{\partial P}{\partial t} + H \cdot \tfrac{\partial M}{\partial t} \quad \text{(per unit volume per unit time)}. \tag{12}$$

Once again, energy conservation is guaranteed by Eq.(9), which is a direct and rigorous consequence of Maxwell's equations, irrespective of how one might interpret the various terms of the equation.

It is noteworthy that the commonly used Poynting vector $S = E \times H$ [60-62] is the one derived in the Einstein-Laub formalism. The Poynting vector $S_L = \mu_0^{-1} E \times B$ associated with the Lorentz interpretation (and preferred by some authors [63,64]) has been criticized on the grounds that it does not maintain the continuity of EM energy flux across the boundary between two adjacent media [61]. The simplest example is provided by a plane EM wave arriving from free space at the flat surface of a semi-infinite magnetic dielectric at normal incidence. The boundary conditions associated with Maxwell's equations dictate the continuity of the $E$ and $H$ components that are parallel to the surface of the medium. Thus, at the entrance facet, the flux of energy associated with $S_L$ exhibits a discontinuity whenever the tangential $B$ field happens to be discontinuous. Proponents of the Lorentz formalism do not dispute this fact, but invoke the existence of a hidden energy flux at the rate of $\mu_0^{-1} M \times E$ that accounts for the discrepancy [8,58]. Be it as it may, since the hidden energy flux is not an observable, one may be forgiven for preferring the formalism that avoids the use of hidden entities.

**5. Electromagnetic force and momentum**. In the Lorentz formalism, all material media are represented by charge and current densities. Generalizing the Lorentz force law $f = q(E + V \times B)$, which is the force exerted on a point-charge $q$ moving with velocity $V$ in the EM fields $E$ and $B$, the force-density which is compatible with the interpretation of Maxwell's equations in accordance with Eqs.(2) may be written as follows [60]:

$$F_L(r,t) = (\rho_{\text{free}} - \nabla \cdot P)E + (J_{\text{free}} + \tfrac{\partial P}{\partial t} + \mu_o^{-1}\nabla \times M) \times B. \tag{13}$$

Substitution for the total charge and current densities from Eqs.(2a) and (2b) into Eq.(13), followed by standard manipulations, yields

$$\overleftrightarrow{\nabla} \cdot \left[\tfrac{1}{2}(\varepsilon_o E \cdot E + \mu_o^{-1} B \cdot B)\overleftrightarrow{I} - \varepsilon_o EE - \mu_o^{-1} BB\right] + \tfrac{\partial}{\partial t}(\varepsilon_o E \times B) + F_L(r,t) = 0. \tag{14}$$

In the above equation, $\overleftrightarrow{I}$ is the identity tensor and the bracketed entity is the Maxwell stress tensor $\overleftrightarrow{\mathcal{T}}(r,t)$. Thus the EM momentum-density in the Lorentz formalism (sometimes referred to as the Livens momentum [65]) is $G(r,t) = \varepsilon_o E \times B = S_L/c^2$. According to Eq.(14), the EM momentum entering through the closed surface of a given volume is equal to the change in the EM momentum stored within that volume plus the mechanical momentum transferred to the material media located inside the volume. The Lorentz force density $F_L(r,t)$ is simply a measure of the transfer rate of momentum from the fields to the material media (or vice versa).

In the Einstein-Laub formalism, the force density, which has contributions from the $E$ and $H$ fields acting on the sources ($\rho_{\text{free}}, J_{\text{free}}, P, M$), is written [3]

$$F_{EL}(r,t) = \rho_{\text{free}} E + J_{\text{free}} \times \mu_o H + (P \cdot \nabla)E + \tfrac{\partial P}{\partial t} \times \mu_o H + (M \cdot \nabla)H - \tfrac{\partial M}{\partial t} \times \varepsilon_o E. \tag{15}$$

Substitution from Eqs.(3) into the above equation, followed by standard algebraic manipulations yields

$$\overleftrightarrow{\nabla} \cdot \left[\tfrac{1}{2}(\varepsilon_o E \cdot E + \mu_o H \cdot H)\overleftrightarrow{I} - DE - BH\right] + \tfrac{\partial}{\partial t}(E \times H/c^2) + F_{EL}(r,t) = 0. \tag{16}$$

The bracketed entity on the right-hand-side of the above equation is the Einstein-Laub stress tensor $\overleftrightarrow{\mathcal{T}}_{EL}(r,t)$. Thus the EM momentum-density in the Einstein-Laub formulation (generally known as the Abraham momentum) is $G(r,t) = E \times H/c^2 = S_{EL}/c^2$. According to Eq.(16), the EM momentum entering through the closed surface of a given volume is equal to the change in the EM momentum stored within that volume plus the mechanical momentum transferred to the material media located inside the volume. The Einstein-Laub force density $F_{EL}(r,t)$ is thus a measure of the transfer rate of momentum from the fields to the media (or vice versa).

Note that the stress tensors of Lorentz and Einstein-Laub, when evaluated in the free-space region surrounding an isolated object, are exactly the same. This means that, in steady-state situations where the enclosed EM momentum does not vary with time, the force exerted on an isolated object in accordance with the Lorentz law is precisely the same as that calculated using the Einstein-Laub formula. Even in situations which depart from the steady-state, the actual force exerted on an isolated object remains the same in the two formulations. Here the difference between the EM momentum densities of Lorentz ($\varepsilon_o E \times B$) and Einstein-Laub ($E \times H/c^2$), namely, $\varepsilon_o E \times M$, accounts only for the hidden mechanical momentum inside magnetic dipoles [8,51-59]. Since hidden momentum has no observable effects on the force and torque exerted on material bodies [29], the difference in the EM momenta in the two formulations cannot have any physical consequences.

It is remarkable that Einstein and Laub proposed their force-density formula, Eq.(15), nearly six decades before Shockley discovered the lack of momentum balance in certain EM systems containing magnetic materials [37-39]. The concept of hidden momentum proposed by Shockley accounts for the momentum imbalance in EM systems that acquire mechanical momentum at a different rate than that dictated by the exerted Lorentz force. Had Shockley used the Einstein-Laub force instead, he would have found perfect balance and no need for hidden momentum.

**6. Electromagnetic torque and angular momentum**. The torque and angular momentum densities in the Lorentz formalism may be determined by cross-multiplying the position vector $\boldsymbol{r}$ into Eq.(14). We will have

$$\boldsymbol{r} \times \frac{\partial}{\partial x}[\tfrac{1}{2}(\varepsilon_0 \boldsymbol{E} \cdot \boldsymbol{E} + \mu_0^{-1}\boldsymbol{B} \cdot \boldsymbol{B})\hat{\boldsymbol{x}} - \varepsilon_0 E_x \boldsymbol{E} - \mu_0^{-1} B_x \boldsymbol{B}]$$

$$+\boldsymbol{r} \times \frac{\partial}{\partial y}[\tfrac{1}{2}(\varepsilon_0 \boldsymbol{E} \cdot \boldsymbol{E} + \mu_0^{-1}\boldsymbol{B} \cdot \boldsymbol{B})\hat{\boldsymbol{y}} - \varepsilon_0 E_y \boldsymbol{E} - \mu_0^{-1} B_y \boldsymbol{B}]$$

$$+\boldsymbol{r} \times \frac{\partial}{\partial z}[\tfrac{1}{2}(\varepsilon_0 \boldsymbol{E} \cdot \boldsymbol{E} + \mu_0^{-1}\boldsymbol{B} \cdot \boldsymbol{B})\hat{\boldsymbol{z}} - \varepsilon_0 E_z \boldsymbol{E} - \mu_0^{-1} B_z \boldsymbol{B}]$$

$$+\boldsymbol{r} \times \frac{\partial}{\partial t}(\varepsilon_0 \boldsymbol{E} \times \boldsymbol{B}) + \boldsymbol{r} \times \boldsymbol{F}_L(\boldsymbol{r}, t) = 0. \tag{17}$$

On the left-hand-side of the above equation, the last term is the Lorentz torque density and the penultimate term is the time-rate-of-change of the EM angular momentum density. As for the remaining terms, one can move $\boldsymbol{r} \times$ inside the differential operators, as is readily verified by simple algebraic differentiation of the resulting expressions. We find

$$\frac{\partial}{\partial x}\{\boldsymbol{r} \times [\tfrac{1}{2}(\varepsilon_0 \boldsymbol{E} \cdot \boldsymbol{E} + \mu_0^{-1}\boldsymbol{B} \cdot \boldsymbol{B})\hat{\boldsymbol{x}} - \varepsilon_0 E_x \boldsymbol{E} - \mu_0^{-1} B_x \boldsymbol{B}]\}$$

$$+\frac{\partial}{\partial y}\{\boldsymbol{r} \times [\tfrac{1}{2}(\varepsilon_0 \boldsymbol{E} \cdot \boldsymbol{E} + \mu_0^{-1}\boldsymbol{B} \cdot \boldsymbol{B})\hat{\boldsymbol{y}} - \varepsilon_0 E_y \boldsymbol{E} - \mu_0^{-1} B_y \boldsymbol{B}]\}$$

$$+\frac{\partial}{\partial z}\{\boldsymbol{r} \times [\tfrac{1}{2}(\varepsilon_0 \boldsymbol{E} \cdot \boldsymbol{E} + \mu_0^{-1}\boldsymbol{B} \cdot \boldsymbol{B})\hat{\boldsymbol{z}} - \varepsilon_0 E_z \boldsymbol{E} - \mu_0^{-1} B_z \boldsymbol{B}]\}$$

$$+\frac{\partial}{\partial t}(\boldsymbol{r} \times \boldsymbol{S}_L/c^2) + \boldsymbol{T}_L(\boldsymbol{r}, t) = 0. \tag{18}$$

The first three terms in Eq.(18) form the divergence of a 2$^{\text{nd}}$ rank tensor, thus confirming the conservation of angular momentum. The EM torque and angular momentum densities in the Lorentz formulation are thus given by

$$\boldsymbol{T}_L(\boldsymbol{r}, t) = \boldsymbol{r} \times \boldsymbol{F}_L(\boldsymbol{r}, t). \tag{19}$$

$$\boldsymbol{\mathcal{L}}_L(\boldsymbol{r}, t) = \boldsymbol{r} \times \boldsymbol{S}_L/c^2. \tag{20}$$

A similar procedure can be carried out within the Einstein-Laub formalism, but the end result, somewhat unexpectedly, turns out to be different. Cross-multiplication of both sides of Eq.(16) into $\boldsymbol{r}$ yields

$$\boldsymbol{r} \times \frac{\partial}{\partial x}[\tfrac{1}{2}(\varepsilon_0 \boldsymbol{E} \cdot \boldsymbol{E} + \mu_0 \boldsymbol{H} \cdot \boldsymbol{H})\hat{\boldsymbol{x}} - D_x \boldsymbol{E} - B_x \boldsymbol{H}]$$

$$+\boldsymbol{r} \times \frac{\partial}{\partial y}[\tfrac{1}{2}(\varepsilon_0 \boldsymbol{E} \cdot \boldsymbol{E} + \mu_0 \boldsymbol{H} \cdot \boldsymbol{H})\hat{\boldsymbol{y}} - D_y \boldsymbol{E} - B_y \boldsymbol{H}]$$

$$+\mathbf{r} \times \frac{\partial}{\partial z}[½(\varepsilon_o \mathbf{E} \cdot \mathbf{E} + \mu_o \mathbf{H} \cdot \mathbf{H})\hat{\mathbf{z}} - D_z \mathbf{E} - B_z \mathbf{H}]$$

$$+\mathbf{r} \times \frac{\partial}{\partial t}(\mathbf{E} \times \mathbf{H}/c^2) + \mathbf{r} \times \mathbf{F}_{EL}(\mathbf{r}, t) = 0. \tag{21}$$

It is not admissible, however, to move $\mathbf{r} \times$ inside the first three differential operators on the left-hand side of Eq.(21) without introducing certain additional terms. Considering that $\partial \mathbf{r}/\partial x = \hat{\mathbf{x}}$, $\partial \mathbf{r}/\partial y = \hat{\mathbf{y}}$, and $\partial \mathbf{r}/\partial z = \hat{\mathbf{z}}$, the requisite additional terms are going to be

$$\hat{\mathbf{x}} \times [½(\varepsilon_o \mathbf{E} \cdot \mathbf{E} + \mu_o \mathbf{H} \cdot \mathbf{H})\hat{\mathbf{x}} - D_x \mathbf{E} - B_x \mathbf{H}]$$
$$+\hat{\mathbf{y}} \times [½(\varepsilon_o \mathbf{E} \cdot \mathbf{E} + \mu_o \mathbf{H} \cdot \mathbf{H})\hat{\mathbf{y}} - D_y \mathbf{E} - B_y \mathbf{H}]$$
$$+\hat{\mathbf{z}} \times [½(\varepsilon_o \mathbf{E} \cdot \mathbf{E} + \mu_o \mathbf{H} \cdot \mathbf{H})\hat{\mathbf{z}} - D_z \mathbf{E} - B_z \mathbf{H}]$$
$$= -\mathbf{D} \times \mathbf{E} - \mathbf{B} \times \mathbf{H} = -\mathbf{P} \times \mathbf{E} - \mathbf{M} \times \mathbf{H}. \tag{22}$$

We may thus rewrite Eq.(21) as follows:

$$\frac{\partial}{\partial x}\{\mathbf{r} \times [½(\varepsilon_o \mathbf{E} \cdot \mathbf{E} + \mu_o \mathbf{H} \cdot \mathbf{H})\hat{\mathbf{x}} - D_x \mathbf{E} - B_x \mathbf{H}]\}$$
$$+\frac{\partial}{\partial y}\{\mathbf{r} \times [½(\varepsilon_o \mathbf{E} \cdot \mathbf{E} + \mu_o \mathbf{H} \cdot \mathbf{H})\hat{\mathbf{y}} - D_y \mathbf{E} - B_y \mathbf{H}]\}$$
$$+\frac{\partial}{\partial z}\{\mathbf{r} \times [½(\varepsilon_o \mathbf{E} \cdot \mathbf{E} + \mu_o \mathbf{H} \cdot \mathbf{H})\hat{\mathbf{z}} - D_z \mathbf{E} - B_z \mathbf{H}]\}$$
$$+\frac{\partial}{\partial t}(\mathbf{r} \times \mathbf{S}_{EL}/c^2) + \mathbf{r} \times \mathbf{F}_{EL}(\mathbf{r}, t) + \mathbf{P} \times \mathbf{E} + \mathbf{M} \times \mathbf{H} = 0. \tag{23}$$

Once again, the first three terms of Eq.(23) form the divergence of a 2$^{nd}$ rank tensor, thus confirming the conservation of angular momentum, while the remaining terms yield expressions for the EM torque and angular momentum densities, as follows:

$$\mathbf{T}_{EL}(\mathbf{r}, t) = \mathbf{r} \times \mathbf{F}_{EL} + \mathbf{P} \times \mathbf{E} + \mathbf{M} \times \mathbf{H}. \tag{24}$$

$$\boldsymbol{\mathcal{L}}_{EL}(\mathbf{r}, t) = \mathbf{r} \times \mathbf{S}_{EL}/c^2. \tag{25}$$

In their original paper [3], Einstein and Laub mentioned the need for the inclusion of $\mathbf{P} \times \mathbf{E}$ and $\mathbf{M} \times \mathbf{H}$ terms in the torque-density expression only briefly and with specific reference to anisotropic bodies. Of course, in linear, isotropic, non-absorbing media, where $\mathbf{P}$ is parallel to $\mathbf{E}$ and $\mathbf{M}$ is parallel to $\mathbf{H}$, both cross-products vanish. However, in more general circumstances, the torque expression *must* include these additional terms. The above derivation of Eq.(23) should make it clear that the EM torque-density of Eq.(24) has general validity, and that conservation of angular momentum demands the addition of $\mathbf{P} \times \mathbf{E}$ and $\mathbf{M} \times \mathbf{H}$ terms to $\mathbf{r} \times \mathbf{F}_{EL}$.

As was the case with the EM force discussed in the preceding section, the effective EM torque exerted on an isolated object always turns out to be the same in the Lorentz and Einstein-Laub formulations; any differences between the two approaches can be reconciled by subtracting the contributions of hidden angular momentum from the Lorentz torque [22,29].

**7. Energy gain and loss by a point-particle**. Consider a point-mass $\mathfrak{m}_o$ moving arbitrarily in free space, its position at time *t* being $[x_p(t), y_p(t), z_p(t)]$. Denoting the instantaneous velocity of the particle by $\mathbf{V}_p(t) = x'_p(t)\hat{\mathbf{x}} + y'_p(t)\hat{\mathbf{y}} + z'_p(t)\hat{\mathbf{z}}$, its mass-density by $m(\mathbf{r}, t)$, its momentum-density by $\wp(\mathbf{r}, t)$, and its energy-density by $\mathcal{E}(\mathbf{r}, t)$, we will have

$$m(\mathbf{r},t) = \mathfrak{m}_o \delta(x - x_p)\delta(y - y_p)\delta(z - z_p), \tag{26}$$

$$\boldsymbol{p}(\mathbf{r},t) = \mathfrak{m}_o \gamma(t) \boldsymbol{V}_p(t) \delta(x - x_p)\delta(y - y_p)\delta(z - z_p), \tag{27}$$

$$\mathcal{E}(\mathbf{r},t) = \mathfrak{m}_o c^2 \gamma(t) \delta(x - x_p)\delta(y - y_p)\delta(z - z_p). \tag{28}$$

In the above equations, $\delta(\cdot)$ is Dirac's delta-function and $\gamma(t) = 1/\sqrt{1 - \boldsymbol{V}_p(t) \cdot \boldsymbol{V}_p(t)/c^2}$ is the Lorentz factor. The time-rate-of-change of the momentum-density is thus given by

$$\frac{\partial}{\partial t}\boldsymbol{p}(\mathbf{r},t) = \mathfrak{m}_o[\gamma'(t)\boldsymbol{V}_p(t) + \gamma(t)\boldsymbol{V}_p'(t)]\delta(x - x_p)\delta(y - y_p)\delta(z - z_p)$$
$$-\mathfrak{m}_o \gamma(t)\boldsymbol{V}_p(t)[V_{px}\delta'(x - x_p)\delta(y - y_p)\delta(z - z_p)$$
$$+ V_{py}\delta(x - x_p)\delta'(y - y_p)\delta(z - z_p)$$
$$+ V_{pz}\delta(x - x_p)\delta(y - y_p)\delta'(z - z_p)]. \tag{29}$$

Integration over the volume of the particle eliminates the last three terms on the right-hand side of Eq.(29). Considering that $\gamma'(t) = \gamma^3 \boldsymbol{V}_p \cdot \boldsymbol{V}_p'/c^2$, we find the time-rate-of-change of the particle's momentum, which is equal to the external force acting on the particle, to be

$$\boldsymbol{F}_{\text{ext}}(t) = d\boldsymbol{\mathcal{P}}(t)/dt = \gamma \mathfrak{m}_o [\gamma^2(\boldsymbol{V}_p \cdot \boldsymbol{V}_p')\boldsymbol{V}_p/c^2 + \boldsymbol{V}_p']. \tag{30}$$

The rate at which energy is injected into (or extracted from) the particle is now found to be

$$\boldsymbol{F}_{\text{ext}}(t) \cdot \boldsymbol{V}_p(t) = \gamma^3 \mathfrak{m}_o \boldsymbol{V}_p \cdot \boldsymbol{V}_p'. \tag{31}$$

In similar fashion, one can evaluate the particle's time-rate-of-change of energy as follows:

$$\iiint_{-\infty}^{\infty} \frac{\partial \mathcal{E}(\mathbf{r},t)}{\partial t} dx dy dz = \gamma'(t) \mathfrak{m}_o c^2 = \gamma^3 \mathfrak{m}_o \boldsymbol{V}_p \cdot \boldsymbol{V}_p'. \tag{32}$$

As expected, Eqs.(31) and (32) confirm that the particle's rate-of-change of energy precisely equals the dot-product of the external force $\boldsymbol{F}_{\text{ext}}(t)$ and the particle's velocity $\boldsymbol{V}_p(t)$. In the special case when a point-charge $q$ travels within an EM field, the energy exchange rate between the field and the particle will be

$$\frac{d}{dt}\mathcal{E}(t) = \boldsymbol{F}_{\text{ext}}(t) \cdot \boldsymbol{V}_p(t) = q(\boldsymbol{E} + \boldsymbol{V}_p \times \boldsymbol{B}) \cdot \boldsymbol{V}_p = q\boldsymbol{E} \cdot \boldsymbol{V}_p. \tag{33}$$

Thus the charged particle either absorbs energy from the field or injects energy into the field, depending on the sign of $q\boldsymbol{E} \cdot \boldsymbol{V}_p$.

The above arguments fail if the rest-mass $\mathfrak{m}_o$ of the particle happens to be time-dependent. Following the same procedure as outlined above, it is not difficult to show that, in general,

$$\frac{d(\mathfrak{m}_o c^2)}{dt} = \gamma\left[\frac{d\mathcal{E}(t)}{dt} - \boldsymbol{F}_{\text{ext}}(t) \cdot \boldsymbol{V}_p(t)\right]. \tag{34}$$

A detailed derivation of Eq.(34) is given in Appendix A. Whenever $d\mathcal{E}(t)/dt$ associated with a particle happens to differ from $\boldsymbol{F}_{\text{ext}}(t) \cdot \boldsymbol{V}_p(t)$, it must be concluded that the particle's rest-mass $\mathfrak{m}_o$ varies with time in accordance with Eq.(34). For a particle at rest, $\boldsymbol{V}_p = 0$ and $\gamma = 1$; Eq.(34) then shows that the entire EM energy gets incorporated into the rest-mass $\mathfrak{m}_o$. In other words, in the absence of moving material bodies, the laws of EM force and EM energy are decoupled from each other. Once the particle has a nonzero velocity, however, Eq.(34) relates the energy and momentum exchanged between the particle and the fields to each other and to the particle's rest-mass $m_o$. While some of the exchanged energy accounts for the variation in the particle's kinetic energy, the remaining part (if any) must get incorporated into its rest-mass. The logic here is that certain fractions of the energy and momentum have disappeared from the field and entered the particle (or vice versa). The particle has rest-mass $m_o$, velocity $V_p$, linear momentum $\gamma m_o V_p$, and total energy $\gamma m_o c^2$. Taking into account the equivalence of mass and energy, the only way to enforce consistency among these entities is to accept the prescription of Eq. (34) for the time-rate-of-change of the rest-mass.

**8. Point-charge traveling in a constant electric field**. In this first example, we demonstrate that a charged particle, acted upon by an external $E$ field, gains or loses kinetic energy, while its rest-mass remains intact. There is no need here to distinguish between the Lorentz and Einstein-Laub formulations, since, in the absence of $P$ and $M$, the two theories are identical.

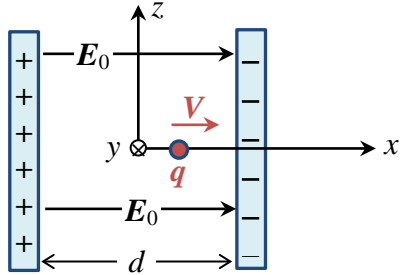

**Fig. 1**. A point-charge $q$ moves at constant velocity $V$ along the $x$-axis. The uniformly-charged parallel plates produce a constant and uniform electric field $E_0 \hat{x}$ in the region between the plates. The plates being non-conductive, their uniform charge distribution is not altered by the point-charge.

Let a point-particle of charge $q$ and (large) mass $m_o$ move at constant velocity $V$ along the $x$-axis. A pair of uniformly-charged parallel plates produces a constant and uniform electric field $E_0$ along the $x$-axis, as shown in Fig. 1. The mass $m_o$ is large enough that the force exerted by the $E$-field on the particle does not produce any significant acceleration, so that the velocity $V$ remains essentially constant. The $E$ and $H$ fields of the moving particle can be found by straightforward Lorentz transformation from the particle's rest frame, as follows:

$$\boldsymbol{E}(\boldsymbol{r},t) = \frac{\gamma q[(x-Vt)\hat{x}+y\hat{y}+z\hat{z}]}{4\pi\varepsilon_o[\gamma^2(x-Vt)^2+y^2+z^2]^{3/2}}, \tag{35a}$$

$$\boldsymbol{H}(\boldsymbol{r},t) = \frac{\gamma qV(-z\hat{y}+y\hat{z})}{4\pi[\gamma^2(x-Vt)^2+y^2+z^2]^{3/2}}. \tag{35b}$$

The energy-densities of $E_y, E_z, H_y, H_z$, when integrated over all space (both inside and outside the cavity formed by the parallel plates), turn out to be independent of the time $t$. The energy-density of $E_x$, however, is mixed with that of the uniform $E$-field $E_0 \hat{x}$ within the cavity, requiring the integration of the cross-term $\varepsilon_o E_0 E_x$ over the volume of the cavity. The left-right symmetry with respect to the particle position allows a reduction of the integration range, as follows:

$$\int_{-\frac{1}{2}d}^{-\frac{1}{2}d+2Vt}\iint_{-\infty}^{\infty}\varepsilon_o E_0 E_x(\boldsymbol{r},t)\mathrm{d}x\mathrm{d}y\mathrm{d}z = \frac{qE_0}{4\pi}\int_{-\frac{1}{2}d}^{-\frac{1}{2}d+2Vt}\iint_{-\infty}^{\infty}\frac{\gamma(x-Vt)}{[\gamma^2(x-Vt)^2+y^2+z^2]^{3/2}}\mathrm{d}x\mathrm{d}y\mathrm{d}z$$

$$= \frac{qE_0}{4\pi\gamma}\int_{-\gamma(\frac{1}{2}d+Vt)}^{-\gamma(\frac{1}{2}d-Vt)}\iint_{-\infty}^{\infty}\frac{x}{(x^2+y^2+z^2)^{3/2}}\mathrm{d}x\mathrm{d}y\mathrm{d}z$$

$$= -qE_0Vt. \tag{36}$$

The time-rate of energy loss/gain by the field is therefore equal to $-qE_0V$, which is precisely equal in magnitude and opposite in sign to the time-rate of energy gain/loss by the moving particle in accordance with Eq.(33). (Note that the finite width $d$ of the cavity is an important feature of the above calculation, even though $d$ itself does not appear in the final result. In the absence of such a cavity, if one simply assumes the existence of a uniform $E$-field throughout the entire space, the total field energy would remain constant as the particle gains kinetic energy while moving forward, thus violating the principle of conservation of energy.)

In summary, the charged particle gains kinetic energy at precisely the same rate as the EM field loses the energy stored in its $E$-field. The right-hand side of Eq.(34) thus vanishes, confirming that the rest-mass $\mathfrak{m}_o$ of the point-charge remains constant as it travels within the cavity depicted in Fig.1.

**9. Electric point-dipole traveling in the $E$-field of a charged wire**. In this second example, a permanent electric dipole exchanges energy and momentum with an inhomogeneous external $E$ field. Both the kinetic energy and the rest-mass of the dipole vary as a result of its interaction with the $E$ field. The Lorentz and Einstein-Laub formalisms exhibit no substantive differences in the present example, neither in the intermediate steps nor in the final result.

Figure 2(a) shows a uniformly-charged, infinitely long wire of charge-density $\lambda_0$ [coulomb/meter] parallel to the $z$-axis. The wire crosses the $xy$-plane at $(x,y) = (-x_0, 0)$. The resulting $E$-field along the $x$-axis is $\boldsymbol{E}(x) = \lambda_0 \hat{\boldsymbol{x}}/[2\pi\varepsilon_o(x+x_0)]$. A massive point-dipole $p_0\hat{\boldsymbol{x}}$ moves at constant velocity $V$ along the $x$-axis; its polarization density (Lorentz-transformed from the rest frame of the particle [60]) is

$$\boldsymbol{P}(\boldsymbol{r},t) = p_0 \delta[\gamma(x-Vt)]\delta(y)\delta(z)\hat{\boldsymbol{x}}. \tag{37}$$

Here, as usual, $\gamma = 1/\sqrt{1-(V/c)^2}$, and $\delta(\cdot)$ is Dirac's delta-function. The right-hand side of the equation may be simplified with the aid of the identity $\delta[\gamma(x-Vt)] = \gamma^{-1}\delta(x-Vt)$.

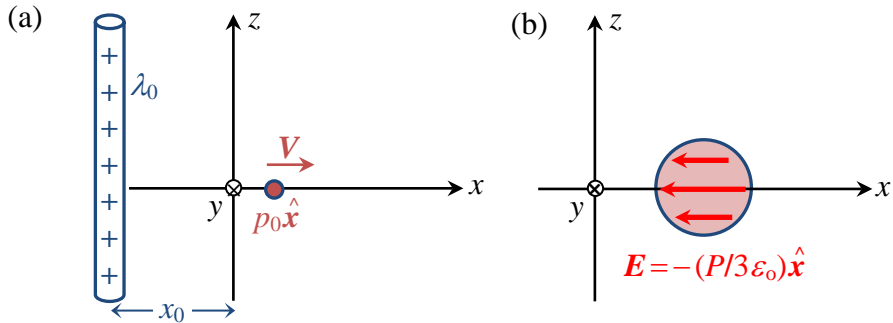

**Fig. 2**. (a) An electric point-dipole $p_0\hat{\boldsymbol{x}}$, traveling at a constant velocity $V$ along the $x$-axis, is acted upon by the $E$ field of a long, straight wire having a linear charge-density $\lambda_0$. The wire, which is parallel to the $z$-axis, crosses the $xy$-plane at $(x,y) = (-x_0, 0)$. (b) Magnified view of the interior of the point-dipole. In its rest frame, the uniformly-polarized sphere of polarization $P\hat{\boldsymbol{x}}$ has a uniform internal field $\boldsymbol{E} = -(P/3\varepsilon_o)\hat{\boldsymbol{x}}$.

There is no relativistically-induced magnetic dipole in this problem, as the electric dipole is oriented along its direction of motion. The dipole does have an internal $E$ field, however, as shown in Fig.2(b). In its rest frame, a uniformly-polarized sphere of polarization $P\hat{\boldsymbol{x}}$ is known to have a uniform internal field $\boldsymbol{E} = -(P/3\varepsilon_o)\hat{\boldsymbol{x}}$ [60]. Therefore, in the $xyz$ frame, the internal field of the dipole is $\boldsymbol{E}(\boldsymbol{r},t) = -(p_0/3\varepsilon_o)\delta[\gamma(x-Vt)]\delta(y)\delta(z)\hat{\boldsymbol{x}}$. The total $E$ field within the dipole is the sum of the external field, produced by the charged wire, and the above internal or self-field. The stored $E$ field energy-density inside the dipole, $\tfrac{1}{2}\varepsilon_o|\boldsymbol{E}_{\text{int}} + \boldsymbol{E}_{\text{ext}}|^2$, thus has three terms of which only the cross-term, $\varepsilon_o \boldsymbol{E}_{\text{int}} \cdot \boldsymbol{E}_{\text{ext}}$, is relevant to the present discussion. Integration over the volume of the dipole thus yields

$$\mathcal{E}(t) = \iiint_{-\infty}^{\infty} \varepsilon_o \boldsymbol{E}_{\text{int}} \cdot \boldsymbol{E}_{\text{ext}} \, \mathrm{d}x\mathrm{d}y\mathrm{d}z$$

$$= -\iiint_{-\infty}^{\infty} \varepsilon_o \frac{\lambda_0}{2\pi\varepsilon_o(x+x_0)} \left(\frac{p_0}{3\varepsilon_o}\right) \delta[\gamma(x-Vt)]\delta(y)\delta(z) \mathrm{d}x\mathrm{d}y\mathrm{d}z = -\frac{\lambda_0 p_0}{6\pi\varepsilon_o \gamma(x_0+Vt)}. \quad (38)$$

The time-rate-of-change of the stored energy inside the dipole is thus $\lambda_0 p_0 V/[6\pi\varepsilon_o \gamma(x_0+Vt)^2]$. To this we must add, in accordance with Eq.(6) (Lorentz) or Eq.(9) (Einstein-Laub), the time-rate of energy exchange between the external $E$ field and the dipole, that is,

$$\iiint_{-\infty}^{\infty} \boldsymbol{E} \cdot \frac{\partial \boldsymbol{P}}{\partial t} \mathrm{d}x\mathrm{d}y\mathrm{d}z = -\iiint_{-\infty}^{\infty} \frac{\lambda_0 p_0 V \delta'(x-Vt)\delta(y)\delta(z)}{2\pi\varepsilon_o \gamma(x+x_0)} \mathrm{d}x\mathrm{d}y\mathrm{d}z = -\frac{\lambda_0 p_0 V}{2\pi\varepsilon_o \gamma(x_0+Vt)^2}. \quad (39)$$

Consequently, the total rate of delivery of EM energy from the external world to the dipole (in both the Lorentz and Einstein-Laub formulations) is

$$\frac{\mathrm{d}\mathcal{E}(t)}{\mathrm{d}t} = -\frac{\lambda_0 p_0 V}{3\pi\varepsilon_o \gamma(x_0+Vt)^2}. \quad (40)$$

Now, in the Einstein-Laub formalism, the force exerted on the dipole is given by

$$\boldsymbol{F}(t) = \iiint_{-\infty}^{\infty} (\boldsymbol{P} \cdot \boldsymbol{\nabla})\boldsymbol{E}(\boldsymbol{r},t)\mathrm{d}x\mathrm{d}y\mathrm{d}z = -\frac{\lambda_0 p_0 \hat{\boldsymbol{x}}}{2\pi\varepsilon_o \gamma(x_0+Vt)^2}. \quad (41)$$

The same expression for the force on the dipole is obtained in the Lorentz formalism as well. In the present problem, since no EM momentum resides within the dipole, the total rate of delivery of momentum from the outside world to the dipole is the force $\boldsymbol{F}(t)$ given by Eq.(41). Invoking Eq.(34), we find

$$\frac{\mathrm{d}(m_o c^2)}{\mathrm{d}t} = \frac{\lambda_0 p_0 V}{6\pi\varepsilon_o(x_0+Vt)^2}. \quad (42)$$

This change in the particle's rest-mass $m_o$ as it moves along the $x$-axis is brought about by interference between the internal $E$ field of the dipole and the external $E$ field produced by the charged wire. In any practical situation, the above change in $m_o$ may or may not be large enough to be measurable. Nevertheless, from a theoretical standpoint, the consistency of the classical laws of force and energy demands that $m_o$ vary with time in accordance with Eq.(42). The present example has also shown that both the Lorentz and Einstein-Laub formulations predict identical behavior for the rest-mass of a permanent electric dipole traveling in an external $E$ field.

The subtle difference in how Eq.(34) is applied to a moving point-charge in Sec.8 and to the moving point-dipole in the present section requires some explanation. The terms appearing on the right-hand-side of Eq.(34), aside from $\gamma$ and $\boldsymbol{V}_p$, which in the present examples are assumed to be essentially constant, are $\mathrm{d}\mathcal{E}/\mathrm{d}t$ and $\boldsymbol{F}_{\text{ext}} = \mathrm{d}\boldsymbol{\mathcal{P}}/\mathrm{d}t$. We thus need to calculate the rates at which EM energy $\mathcal{E}(t)$ and EM momentum $\boldsymbol{\mathcal{P}}(t)$ enter or exit through the surface immediately surrounding the particle. This may be done by integrating the Poynting vector $\boldsymbol{S}(\boldsymbol{r},t)$ and the stress tensor $\overleftrightarrow{\boldsymbol{\mathcal{T}}}(\boldsymbol{r},t)$ over the surrounding surface. The task is substantially simplified, however, if we use Eqs.(6) and (9) (Poynting's theorem) and Eqs.(14) and (16) (momentum conservation law) to replace the surface integrals with integrals over the volume of the particle. (See Appendix B for further discussion.)

The requisite volume integrals often involve the squared amplitude of an individual field, or the squared sum of two field amplitudes, or the dot- and cross-products of two different fields inside the volume of the particle. We are thus left with terms that contain the squared internal field of the dipole, the squared external field, and the cross-term between an internal field and an external field. Of these, only the cross-terms contribute to $d\mathcal{E}/dt$ and $d\mathcal{P}/dt$, as the internal fields are time-independent, while the external fields are too small to survive after being multiplied by the (minuscule) volume of a point-dipole. In contrast, the cross-terms have both the time-dependence of the external field and the "giant" magnitude of the internal field to survive integration over the small volume followed by differentiation with respect to time. The surviving cross-terms then contribute to the time-rate-of-change of energy and momentum associated with a point-dipole. The cross-terms fail to survive, however, in the case of a point-charge.

**10. Magnetic point-dipole traveling in the *H*-field of a current-carrying wire**. Our third example examines a permanent magnetic dipole in the process of exchanging energy and momentum with a static, nonuniform external $H$ field. Both the kinetic energy and the rest-mass of the dipole will be seen to change as a result of this interaction. Although the Lorentz and Einstein-Laub formalisms in the present example exhibit substantial differences in the intermediate steps, the final results turn out to be the same.

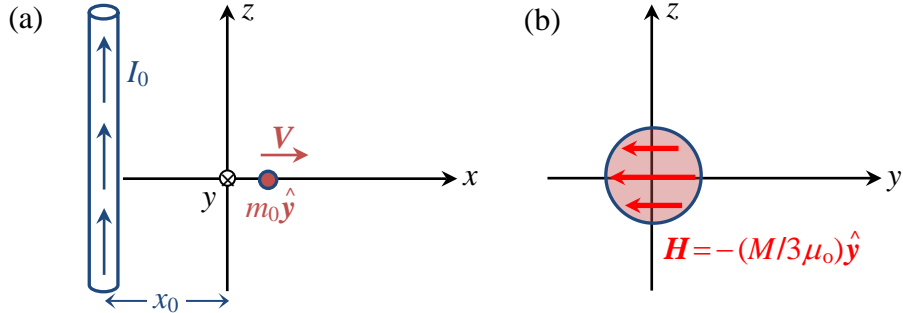

**Fig. 3**. (a) A magnetic point-dipole $m_0\hat{\pmb{y}}$ travels at a constant velocity $V$ along the $x$-axis. The external $H$-field is produced by a long, straight wire carrying a constant current $I_0$. The wire is parallel to the $z$-axis and crosses the $xy$-plane at $(x,y) = (-x_0, 0)$. (b) Magnified view of the interior of the point-dipole. In its rest frame, the uniformly-magnetized sphere of magnetization $M\hat{\pmb{y}}$ has a uniform internal field $\pmb{H} = -(M/3\mu_\mathrm{o})\hat{\pmb{y}}$.

Figure 3(a) shows an infinitely long, straight wire, carrying the constant current $I_0$ parallel to the $z$-axis. The wire crosses the $xy$-plane at $(x,y) = (-x_0, 0)$. In a cylindrical coordinate system centered on the wire, the magnetic field thus produced is

$$\pmb{H}(\pmb{r},t) = (I_0/2\pi r)\hat{\pmb{\phi}}. \tag{43}$$

Let a massive magnetic point-dipole $m_0\hat{\pmb{y}}$, accompanied by a relativistically-induced electric point-dipole $p_0\hat{\pmb{z}}$, travel at the constant velocity $V$ along the $x$-axis. The magnetization and polarization of the dipole-pair may be written as

$$\pmb{M}(\pmb{r},t) = \gamma m_0 \delta[\gamma(x - Vt)]\delta(y)\delta(z)\hat{\pmb{y}}, \tag{44a}$$

$$\pmb{P}(\pmb{r},t) = \gamma\varepsilon_\mathrm{o} m_0 V \delta[\gamma(x - Vt)]\delta(y)\delta(z)\hat{\pmb{z}}. \tag{44b}$$

If we allow a small solid sphere to represent the original point-dipole $\boldsymbol{m}_0$, the internal $H$-field in the dipole's rest-frame will be $\boldsymbol{H}_{\text{int}} = -(m_0/3\mu_o)\delta(x')\delta(y')\delta(z')\widehat{\boldsymbol{y}}'$. A Lorentz transformation to the $xyz$ frame then yields the fields internal to the spherical volume as follows:

$$\boldsymbol{E}_{\text{int}}(\boldsymbol{r},t) = -(2m_0 V/3)\delta(x-Vt)\delta(y)\delta(z)\widehat{\boldsymbol{z}}, \tag{45a}$$

$$\boldsymbol{D}_{\text{int}}(\boldsymbol{r},t) = (\varepsilon_o m_0 V/3)\delta(x-Vt)\delta(y)\delta(z)\widehat{\boldsymbol{z}}, \tag{45b}$$

$$\boldsymbol{B}_{\text{int}}(\boldsymbol{r},t) = (2m_0/3)\delta(x-Vt)\delta(y)\delta(z)\widehat{\boldsymbol{y}}, \tag{45c}$$

$$\boldsymbol{H}_{\text{int}}(\boldsymbol{r},t) = -(m_0/3\mu_o)\delta(x-Vt)\delta(y)\delta(z)\widehat{\boldsymbol{y}}. \tag{45d}$$

It is interesting to note that the internal $E$ and $D$ fields are *not* the same as those expected from a true (as opposed to relativistically-induced) permanent dipole $\boldsymbol{p}_0 = p_0 \widehat{\boldsymbol{z}}$.

In the Einstein-Laub formalism, the force-density on the dipole-pair is evaluated as follows:

$$\boldsymbol{F}_{EL}(\boldsymbol{r},t) = m_0 \delta(x-Vt)\delta(y)\delta(z)\frac{\partial}{\partial y}\frac{I_0[(x+x_0)\widehat{\boldsymbol{y}}-y\widehat{\boldsymbol{x}}]}{2\pi[(x+x_0)^2+y^2]}\bigg|_{y=0}$$

$$-\varepsilon_o m_0 V^2 \delta'(x-Vt)\delta(y)\delta(z)\widehat{\boldsymbol{z}} \times \mu_o I_0 \widehat{\boldsymbol{y}}/[2\pi(x+x_0)]$$

$$= -\frac{m_0 I_0}{2\pi(x+x_0)^2}\delta(x-Vt)\delta(y)\delta(z)\widehat{\boldsymbol{x}} + \frac{m_0 I_0 (V/c)^2}{2\pi(x+x_0)}\delta'(x-Vt)\delta(y)\delta(z)\widehat{\boldsymbol{x}}. \tag{46}$$

Integration over the volume of the dipole yields

$$\boldsymbol{F}_{EL}(t) = -\frac{m_0 I_0 \widehat{\boldsymbol{x}}}{2\pi\gamma^2(x_0+Vt)^2}. \tag{47}$$

The internal EM momentum of the dipole is given by

$$\boldsymbol{\mathcal{P}}_{EL}^{(\text{EM})} = \frac{1}{c^2}\iiint_{-\infty}^{\infty}[-\tfrac{2}{3}m_0 V \delta(x-Vt)\delta(y)\delta(z)\widehat{\boldsymbol{z}}] \times \frac{I_0 \widehat{\boldsymbol{y}}}{2\pi(x+x_0)}\text{d}x\text{d}y\text{d}z = \frac{m_0 I_0 V \widehat{\boldsymbol{x}}}{3\pi c^2(x_0+Vt)}. \tag{48}$$

The time-rate-of-change of the above internal momentum is $-m_0 I_0 V^2 \widehat{\boldsymbol{x}}/[3\pi c^2(x_0+Vt)^2]$, which must be added to $\boldsymbol{F}_{EL}(t)$ of Eq.(47) to yield the total rate-of-change of the momentum transferred from the external world to the dipole-pair. We find

$$\boldsymbol{F}_{EL}(t) + \frac{\text{d}\boldsymbol{\mathcal{P}}_{EL}^{(\text{EM})}}{\text{d}t} = -\frac{m_0 I_0 (1-V^2/3c^2)\widehat{\boldsymbol{x}}}{2\pi(x_0+Vt)^2}. \tag{49}$$

With reference to Eq.(16), note that Eq.(49) yields the *total* rate of transfer of linear momentum from the outside world to the moving dipole-pair. In other words, by adding up the force and the time-rate-of-change of the internal momentum, we have essentially evaluated the integral of the stress tensor $\overset{\leftrightarrow}{\boldsymbol{\mathcal{T}}}_{EL}(\boldsymbol{r},t)$ over a closed surface surrounding the moving dipole-pair. This is what we are going to substitute later on for $\boldsymbol{F}_{\text{ext}}(t)$ in Eq.(34). We follow a similar procedure to calculate the total rate of transfer of energy from the outside world to the dipole-pair. With reference to Eq.(12), the energy exchange rate between the field and the dipole-pair is

$$\frac{\text{d}\mathcal{E}_{EL}^{(\text{exch})}}{\text{d}t} = \iiint_{-\infty}^{\infty}\boldsymbol{H}\cdot\frac{\partial\boldsymbol{M}}{\partial t}\text{d}x\text{d}y\text{d}z = -\frac{m_0 I_0 V}{2\pi(x_0+Vt)^2}. \tag{50}$$

The relevant part of the internal energy of the dipole, produced by interference between $\boldsymbol{H}_{\text{ext}}$ and $\boldsymbol{H}_{\text{int}}$ (i.e., the cross-term), is given by

$$-\iiint_{-\infty}^{\infty}\mu_o\{I_0/[2\pi(x+x_0)]\}(m_0/3\mu_o)\delta(x-Vt)\delta(y)\delta(z)\text{d}x\text{d}y\text{d}z = -\frac{m_0 I_0}{6\pi(x_0+Vt)}. \tag{51}$$

Differentiating the above expression with respect to time yields $m_0 I_0 V/[6\pi(x_0+Vt)^2]$ for the time-rate-of-change of the internal energy of the dipole. Adding this to the energy exchange rate in Eq.(50), we find

$$\frac{d\mathcal{E}_{EL}(t)}{dt} = -\frac{m_0 I_0 V}{3\pi(x_0+Vt)^2}. \tag{52}$$

With reference to Eq.(9), note that Eq.(52) gives the *total* rate of transfer of energy from the outside world to the moving dipole-pair. In other words, by adding the time-rate-of-change of the internally stored energy to the energy exchange rate, we have essentially evaluated the integral of the Poynting vector $\boldsymbol{S}_{EL}(\boldsymbol{r},t)$ over a closed surface surrounding the moving dipole-pair. For the present problem, this constitutes the term $d\mathcal{E}(t)/dt$ to be used in Eq.(34). Substitution from Eqs.(49) and (52) into Eq.(34) thus yields

$$\frac{d(\mathfrak{m}_0 c^2)}{dt} = \frac{m_0 I_0 V}{6\pi\gamma(x_0+Vt)^2}. \tag{53}$$

This time-dependence of the particle's rest-mass as it travels along the *x*-axis is associated with interference between the internal *H*-field of the magnetic dipole and the external *H*-field produced by the long wire depicted in Fig.3.

Next we repeat the same procedures in the Lorentz formalism. The force on the dipole-pair is evaluated as follows:

$$\boldsymbol{F}_L(t) = \iiint_{-\infty}^{\infty}(\mu_0^{-1}\boldsymbol{\nabla}\times\boldsymbol{M} + \partial\boldsymbol{P}/\partial t)\times\boldsymbol{B}\,dxdydz = -\frac{m_0 I_0 \hat{\boldsymbol{x}}}{2\pi\gamma^2(x_0+Vt)^2}. \tag{54}$$

This is precisely the same force as obtained in Eq.(47) using the Einstein-Laub formula. Also, the time-dependent part of the internal EM momentum of the dipole remains the same as that in Eq.(48)—because the contribution of $\varepsilon_0\boldsymbol{E}\times\boldsymbol{M}$ is time-independent. Therefore, Eq.(49) remains the same in the two formalisms.

In the Lorentz formulation, the rate of energy exchange between the field and the dipoles is zero, as this rate is given by $\iiint_{-\infty}^{\infty}\boldsymbol{E}\cdot(\mu_0^{-1}\boldsymbol{\nabla}\times\boldsymbol{M} + \partial\boldsymbol{P}/\partial t)dxdydz$, and the system of Fig.3 lacks an external *E*-field. However, the stored (internal) energy of the dipoles, produced by interference between $\boldsymbol{B}_{\text{ext}}$ and $\boldsymbol{B}_{\text{int}}$ (the cross-term) is now given by

$$\mathcal{E}_L(t) = \iiint_{-\infty}^{\infty}\mu_0^{-1}\left[\frac{\mu_0 I_0}{2\pi(x+x_0)}\right][\tfrac{2}{3}m_0\delta(x-Vt)\delta(y)\delta(z)]dxdydz = \frac{m_0 I_0}{3\pi(x_0+Vt)}. \tag{55}$$

It is seen that the rate of delivery of EM energy from the external world to the dipole-pair, obtained by differentiating Eq.(55) with respect to time, is the same as that given by Eq.(52). Consequently, the time-rate-of-change of the rest-mass $\mathfrak{m}_0$ is the same as that given by Eq.(53). Once again, the predictions of the two formulations are seen to be identical. We emphasize that, quite aside from the question of whether in a practical setting the change of $\mathfrak{m}_0$ would be large enough to be measureable, from a theoretical standpoint, the consistency of the classical laws of force and energy demands that $\mathfrak{m}_0$ vary with time in accordance with Eq.(53). The physical basis for the variation of $\mathfrak{m}_0$ is, of course, the interference between the internal magnetic field of the dipole depicted in Fig.3(b) and the externally applied magnetic field.

**11. Standing wave acting on an electric dipole**. In this example we consider a standing EM wave confined between two perfectly conducting parallel plates located at $z=\pm\tfrac{1}{2}d$, as shown in Fig.4. The *E* and *H* fields of the wave, which is linearly polarized along the *x*-axis, are given by

$$\boldsymbol{E}(\boldsymbol{r},t) = E_0 \sin(k_0 z)\cos(\omega t)\hat{\boldsymbol{x}}, \qquad (56a)$$

$$\boldsymbol{H}(\boldsymbol{r},t) = -(E_0/Z_o)\cos(k_0 z)\sin(\omega t)\hat{\boldsymbol{y}}. \qquad (56b)$$

In the above expressions, $\omega$ is the angular frequency of the monochromatic wave, $k_0 = \omega/c = 2\pi/\lambda$ is the wavenumber, $\lambda$ is the vacuum wavelength, $Z_o = \sqrt{\mu_o/\varepsilon_o}$ is the impedance of free space, and $c = 1/\sqrt{\mu_o\varepsilon_o}$ is the speed of light in vacuum. Since, at the surface of the conducting plates, the $E$-field must vanish, we have $d = m\lambda$, where $m$ is an arbitrary positive integer. The surface-current-density on the inner facets of the conducting plates is, therefore, given by the magnitude of the $H$-field in the immediate vicinity of the plates, namely,

$$\boldsymbol{J}_s = \pm(E_0/Z_o)\sin(\omega t)\hat{\boldsymbol{x}}. \qquad (57)$$

Here the plus and minus signs represent the upper and lower plates (not necessarily in that order), with the upper plate having the plus (minus) sign when $m$ is an odd (even) integer. The stored EM energy (per unit area of the $xy$-plane) is readily found to be

$$\mathcal{E}_{\text{total}} = \int_{-\frac{1}{2}d}^{\frac{1}{2}d} (\tfrac{1}{2}\varepsilon_o \boldsymbol{E}\cdot\boldsymbol{E} + \tfrac{1}{2}\mu_o \boldsymbol{H}\cdot\boldsymbol{H})\mathrm{d}z = \tfrac{1}{4}\varepsilon_o E_0^2 d. \qquad (58)$$

In what follows, we shall find that the total energy of the system remains constant at all times. A fraction of the field energy given in Eq.(58) will be exchanged with a point-dipole that will be introduced into the cavity, but no new energy enters, nor any existing energy leaves the system. When an $E$-field is found to act upon the surface currents $\boldsymbol{J}_s$, or when new currents are induced in the conducting plates, the symmetry of the problem will be such that the integral of $\boldsymbol{E}\cdot\boldsymbol{J}$ over the surface of each plate will remain zero at all times. In other words, no EM energy can enter or exit the system through these plates.

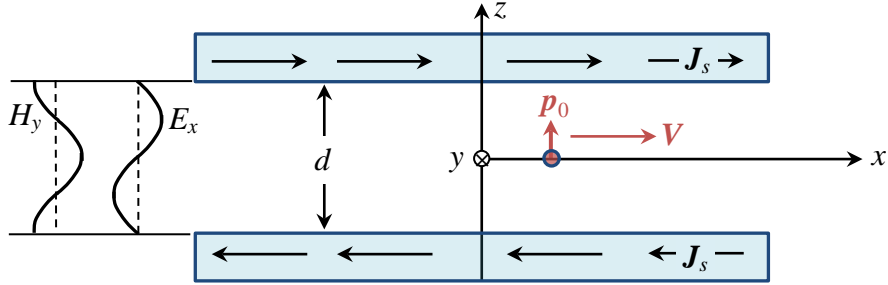

**Fig.4**. A monochromatic standing-wave resides in the space between two perfectly conducting plates. While the $E$-field has a null in the central $xy$-plane, the $H$-field has a peak in that plane. The surface currents $\boldsymbol{J}_s$ sustain the standing wave. An electric point-dipole $\boldsymbol{p}_0 = p_0\hat{\boldsymbol{z}}$ moves along the $x$-axis at a constant velocity $V$, giving rise to a relativistically-induced magnetic dipole $\boldsymbol{m}_0 = m_0\hat{\boldsymbol{y}}$.

Let a permanent electric point-dipole $\boldsymbol{p}_0 = p_0\hat{\boldsymbol{z}}$, whose magnitude and orientation remain fixed in space-time, move at a constant velocity $V$ along the $x$-axis, as shown in Fig.4. The relativistically induced magnetic point-dipole which accompanies $\boldsymbol{p}_0$ will then be $\boldsymbol{m}_0 = \mu_o p_0 V \hat{\boldsymbol{y}}$. The polarization and magnetization associated with this moving dipole-pair are given by

$$\boldsymbol{P}(\boldsymbol{r},t) = \gamma p_0 \delta[\gamma(x - Vt)]\delta(y)\delta(z)\hat{\boldsymbol{z}}, \qquad (59a)$$

$$\boldsymbol{M}(\boldsymbol{r},t) = \gamma\mu_o p_0 V\delta[\gamma(x - Vt)]\delta(y)\delta(z)\hat{\boldsymbol{y}}. \qquad (59b)$$

Allowing a small solid sphere to represent the original point-dipole $\boldsymbol{p}_0$, we find the internal $E$-field in the dipole's rest frame to be $\boldsymbol{E}_{\text{int}} = -(p_0/3\varepsilon_o)\delta(x')\delta(y')\delta(z')\hat{\boldsymbol{z}}'$. Lorentz transformation to the $xyz$ frame then yields the internal fields (i.e., fields within the spherical volume) as follows:

$$\boldsymbol{E}_{\text{int}}(\boldsymbol{r},t) = -(p_0/3\varepsilon_\text{o})\delta(x - Vt)\delta(y)\delta(z)\hat{\boldsymbol{z}}, \tag{60a}$$

$$\boldsymbol{D}_{\text{int}}(\boldsymbol{r},t) = (2p_0/3)\delta(x - Vt)\delta(y)\delta(z)\hat{\boldsymbol{z}}, \tag{60b}$$

$$\boldsymbol{B}_{\text{int}}(\boldsymbol{r},t) = (\mu_\text{o} p_0 V/3)\delta(x - Vt)\delta(y)\delta(z)\hat{\boldsymbol{y}}, \tag{60c}$$

$$\boldsymbol{H}_{\text{int}}(\boldsymbol{r},t) = -(2p_0 V/3)\delta(x - Vt)\delta(y)\delta(z)\hat{\boldsymbol{y}}. \tag{60d}$$

It is interesting to note that the internal $B$ and $H$ fields are *not* the same as those expected from a true (as opposed to relativistically-induced) permanent dipole $\boldsymbol{m}_0 = m_0\hat{\boldsymbol{y}}$.

**11.1. Force and energy in the Einstein-Laub formalism**. The force exerted on the dipole-pair depicted in Fig.4 may be calculated in accordance with the Einstein-Laub theory as follows:

$$\begin{aligned}\boldsymbol{F}_{EL}(\boldsymbol{r},t) &= (\boldsymbol{P} \cdot \boldsymbol{\nabla})\boldsymbol{E} + (\partial \boldsymbol{P}/\partial t) \times \mu_\text{o}\boldsymbol{H} + (\boldsymbol{M} \cdot \boldsymbol{\nabla})\boldsymbol{H} - (\partial \boldsymbol{M}/\partial t) \times \varepsilon_\text{o}\boldsymbol{E} \\ &= p_0\delta(x - Vt)\delta(y)\delta(z)E_0 k_0 \cos(k_0 z)\cos(\omega t)\hat{\boldsymbol{x}} \\ &\quad -(V/c)p_0\delta'(x - Vt)\delta(y)\delta(z)E_0 \cos(k_0 z)\sin(\omega t)\hat{\boldsymbol{x}} \\ &\quad -(V^2/c^2)p_0\delta'(x - Vt)\delta(y)\delta(z)E_0 \sin(k_0 z)\cos(\omega t)\hat{\boldsymbol{z}}. \end{aligned} \tag{61}$$

Integration over the dipole volume yields the following total force on the moving dipole-pair:

$$\boldsymbol{F}_{EL}(t) = p_0 k_0 E_0 \cos(\omega t)\hat{\boldsymbol{x}}. \tag{62}$$

There is also the internal EM momentum of the dipole, whose density is given by

$$\boldsymbol{p}_{EL}^{(\text{EM})}(\boldsymbol{r},t) = \boldsymbol{E}_{\text{int}} \times \boldsymbol{H}/c^2 = -(p_0/3c)\delta(x - Vt)\delta(y)\delta(z)E_0 \cos(k_0 z)\sin(\omega t)\hat{\boldsymbol{x}}. \tag{63}$$

Differentiation with respect to time, followed by integration over the volume of the dipole, yields the corresponding force experienced by the dipole as

$$\boldsymbol{F}_{\text{int}}(t) = -\tfrac{1}{3}p_0 k_0 E_0 \cos(\omega t)\hat{\boldsymbol{x}}. \tag{64}$$

Adding this result to the force expression of Eq.(62), one finds the net force acting on the dipole to be $\tfrac{2}{3}\boldsymbol{F}_{EL}(t)$. Next, we calculate the rate of exchange of energy between the fields and the dipoles:

$$\frac{\partial \mathcal{E}_{EL}}{\partial t} = \boldsymbol{E} \cdot \frac{\partial \boldsymbol{P}}{\partial t} + \boldsymbol{H} \cdot \frac{\partial \boldsymbol{M}}{\partial t} = (V^2/c)p_0 E_0 \cos(k_0 z)\sin(\omega t)\delta'(x - Vt)\delta(y)\delta(z). \tag{65}$$

The above expression integrates to zero over the volume of the dipoles. Despite the fact that the EM field continually exerts a force on the dipole pair, no energy appears to have been exchanged between the field and the dipoles. However, the internal energy-density of the dipole pair, may be evaluated as follows:

$$\begin{aligned}\mathcal{E}_{EL}^{(\text{int})}(\boldsymbol{r},t) &= \tfrac{1}{2}\varepsilon_\text{o} E^2 + \tfrac{1}{2}\mu_\text{o} H^2 \\ &= \tfrac{1}{2}\varepsilon_\text{o}\left[\tfrac{1}{3}p_0\delta(x - Vt)\delta(y)\delta(z)\right]^2 \\ &\quad + \tfrac{1}{2}\mu_\text{o}\left[\tfrac{2}{3}Vp_0\delta(x - Vt)\delta(y)\delta(z) + (E_0/Z_\text{o})\cos(k_0 z)\sin(\omega t)\right]^2. \end{aligned} \tag{66}$$

Emphasizing the crucial contribution to the above expression arising from interference between the internal and external $H$ fields, we may ignore the constant or insignificant terms and rewrite Eq.(66) as

$$\mathcal{E}_{EL}^{(\text{int})}(\mathbf{r},t) = \tfrac{2}{3}(V/c)E_0 p_0 \delta(x-Vt)\delta(y)\delta(z)\cos(k_0 z)\sin(\omega t) + \text{Inconsequential terms.} \quad (67)$$

Differentiation with respect to time, followed by integration over the volume of the dipole-pair now yields

$$\iiint_{-\infty}^{\infty} \frac{\partial \mathcal{E}_{EL}^{(\text{int})}}{\partial t} dxdydz = \tfrac{2}{3} V p_0 k_0 E_0 \cos(\omega t). \quad (68)$$

This is the same as the rate of injection of energy into the dipole-pair by the net force $\tfrac{2}{3}\mathbf{F}_{EL}(t)$ obtained earlier. Invoking Eq.(34), we conclude that the rest-mass $\mathfrak{m}_o$ of the point-dipole in the present example remains constant. Thus, when the fields inside a dipole carry their own momentum, it is possible for the force and energy laws to remain consistent with each other without requiring a concomitant change in the rest mass $\mathfrak{m}_o$ of the dipole.

**11.2. Force and energy in the Lorentz formalism**. Next, we compute the force exerted on the dipole-pair in the system of Fig.4 using the Lorentz formulation, namely,

$$\begin{aligned}\mathbf{F}_L(\mathbf{r},t) &= -(\boldsymbol{\nabla}\cdot\mathbf{P})\mathbf{E} + (\partial\mathbf{P}/\partial t + \mu_o^{-1}\boldsymbol{\nabla}\times\mathbf{M})\times\mathbf{B} \\ &= -p_0\delta(x-Vt)\delta(y)\delta'(z)E_0\sin(k_0 z)\cos(\omega t)\,\hat{\mathbf{x}} \\ &\quad +(V/c)p_0\delta(x-Vt)\delta(y)\delta'(z)E_0\cos(k_0 z)\sin(\omega t)\,\hat{\mathbf{z}}. \end{aligned} \quad (69)$$

Integration over the dipole volume yields the total Lorentz force as

$$\mathbf{F}_L(t) = p_0 k_0 E_0 \cos(\omega t)\,\hat{\mathbf{x}}. \quad (70)$$

This is the same as the Einstein-Laub force experienced by the dipole pair; see Eq.(62). Combining this force with the internal force $\mathbf{F}_{\text{int}}(t)$ of Eq.(64) yields the net force as $\tfrac{2}{3}\mathbf{F}_L(t)$, as before. [Strictly speaking, the internal momentum density of the dipole in the Lorentz formalism is $\varepsilon_o\mathbf{E}_{\text{int}}\times\mathbf{B} = \varepsilon_o\mathbf{E}_{\text{int}}\times(\mu_o\mathbf{H}_{\text{ext}} + \mathbf{B}_{\text{int}})$. However, the contribution of $\mathbf{B}_{\text{int}}$ to the internal momentum is time-independent and can, therefore, be ignored.]

The rate of transfer of energy from the field to the dipoles in the Lorentz formulation is readily found to be

$$\begin{aligned}\frac{\partial \mathcal{E}_L}{\partial t} &= \mathbf{E}\cdot\left(\frac{\partial\mathbf{P}}{\partial t} + \mu_o^{-1}\boldsymbol{\nabla}\times\mathbf{M}\right) \\ &= -V p_0 \delta(x-Vt)\delta(y)\delta'(z)E_0\sin(k_0 z)\cos(\omega t). \end{aligned} \quad (71)$$

Integration over the volume of the dipoles yields

$$\iiint_{-\infty}^{\infty} \frac{\partial \mathcal{E}_L}{\partial t} dxdydz = V p_0 k_0 E_0 \cos(\omega t). \quad (72)$$

Moreover, the internal energy-density of the dipole-pair is given by

$$\begin{aligned}\mathcal{E}_L^{(\text{int})} &= \tfrac{1}{2}\varepsilon_o E^2 + \tfrac{1}{2}\mu_o^{-1} B^2 \\ &= \tfrac{1}{2}\varepsilon_o\left[\tfrac{1}{3}p_0\delta(x-Vt)\delta(y)\delta(z)\right]^2 \\ &\quad + \tfrac{1}{2}\mu_o^{-1}\left[\tfrac{1}{3}V\mu_o p_0\delta(x-Vt)\delta(y)\delta(z) - \mu_o(E_0/Z_o)\cos(k_0 z)\sin(\omega t)\right]^2. \end{aligned} \quad (73)$$

Emphasizing the crucial contribution to the above expression arising from interference between the internal and external $B$ fields, we may once again ignore the constant or insignificant terms and rewrite Eq.(73) as

$$\mathcal{E}_L^{(\text{int})} = -\tfrac{1}{3}(V/c)p_0 E_0 \delta(x - Vt)\delta(y)\delta(z)\cos(k_0 z)\sin(\omega t) + \text{Inconsequential terms.} \tag{74}$$

Differentiation with respect to time, followed by integration over the dipole volume now yields

$$\iiint_{-\infty}^{\infty} \frac{\partial \mathcal{E}_L^{(\text{int})}}{\partial t} \, dx dy dz = -\tfrac{1}{3} V p_0 k_0 E_0 \cos(\omega t). \tag{75}$$

The sum of Eqs.(72) and (75) coincides with Eq.(68). Therefore, the energy transfer rate from the external world to the moving dipole is the same in both formulations. Once again, invoking Eq.(34) in conjunction with the sum of Eqs.(72) and (75) and with the fact that the net force acting on the dipole is ⅔ of $\boldsymbol{F}_L(t)$ of Eq.(70), we find that the rest-mass $\mathfrak{m}_0$ of the point-dipole remains constant.

**12. Standing wave acting on a magnetic dipole**. We now consider a slightly different EM wave between two parallel plates located at $z=\pm\tfrac{1}{2}d$, as shown in Fig.5. The $E$ and $H$ fields of the wave, which is now linearly polarized along the $y$-axis, are given by

$$\boldsymbol{H}(\boldsymbol{r},t) = (E_0/Z_0)\sin(k_0 z)\sin(\omega t)\,\hat{\boldsymbol{x}}, \tag{76a}$$

$$\boldsymbol{E}(\boldsymbol{r},t) = -E_0 \cos(k_0 z)\cos(\omega t)\,\hat{\boldsymbol{y}}. \tag{76b}$$

Since, at the surface of the conducting plates, the $E$-field must vanish, we have $d = (m - \tfrac{1}{2})\lambda$, where $m$ is an arbitrary positive integer. The surface-current-density on the inner facets of the conducting plates is, therefore, given by the magnitude of the $H$-field in the immediate vicinity of the plates, namely,

$$\boldsymbol{J}_s = \pm(E_0/Z_0)\sin(\omega t)\hat{\boldsymbol{y}}. \tag{77}$$

Here the surface currents of both upper and lower plates have the same sign, with the plus (minus) sign corresponding to the case of $m$ being an even (odd) integer. The EM energy per unit area of the $xy$-plane remains the same as that in the system of Fig.4, namely, $\mathcal{E}_{\text{total}} = \tfrac{1}{4}\varepsilon_0 E_0^2 d$.

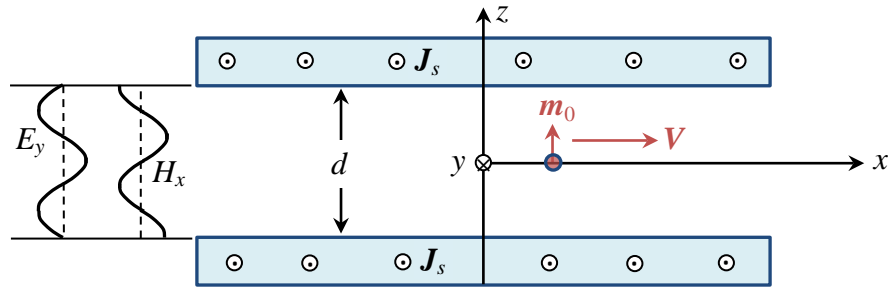

**Fig.5**. A monochromatic standing-wave resides in the space between two perfectly conducting plates. While the $E$-field peaks in the central $xy$-plane, the $H$-field has a null in that plane. The surface currents $\boldsymbol{J}_s$ sustain the standing wave. A magnetic point-dipole $\boldsymbol{m}_0 = m_0 \hat{\boldsymbol{z}}$ travels along the $x$-axis at a constant velocity $V$, giving rise to a relativistically-induced electric dipole $\boldsymbol{p}_0 = p_0 \hat{\boldsymbol{y}}$.

Next, we assume a magnetic point-dipole $\boldsymbol{m}_0 = m_0 \hat{\boldsymbol{z}}$ travels at constant velocity $V$ along the $x$-axis in the $xy$-plane of the system of Fig.5. The induced electric dipole will now be $\boldsymbol{p}_0 = p_0 \hat{\boldsymbol{y}}$, and the magnetization and polarization as functions of the space-time coordinates will be

$$M(r,t) = \gamma m_0 \delta[\gamma(x - Vt)]\delta(y)\delta(z)\hat{z}, \tag{78a}$$

$$P(r,t) = -\gamma\varepsilon_0 m_0 V\delta[\gamma(x - Vt)]\delta(y)\delta(z)\hat{y}. \tag{78b}$$

Allowing a small solid sphere to represent the original point-dipole $m_0$, we find the internal $H$-field in the dipole's rest frame to be $H_{\text{int}} = -(m_0/3\mu_0)\delta(x')\delta(y')\delta(z')\hat{z}'$. A Lorentz transformation to the $xyz$ frame then yields the fields internal to the spherical volume as follows:

$$E_{\text{int}}(r,t) = (2m_0 V/3)\delta(x - Vt)\delta(y)\delta(z)\hat{y}, \tag{79a}$$

$$D_{\text{int}}(r,t) = -(\varepsilon_0 m_0 V/3)\delta(x - Vt)\delta(y)\delta(z)\hat{y}, \tag{79b}$$

$$B_{\text{int}}(r,t) = (2m_0/3)\delta(x - Vt)\delta(y)\delta(z)\hat{z}, \tag{79c}$$

$$H_{\text{int}}(r,t) = -(m_0/3\mu_0)\delta(x - Vt)\delta(y)\delta(z)\hat{z}. \tag{79d}$$

Note, once again, that the internal $E$ and $D$ fields are *not* the same as those expected from a true (as opposed to relativistically-induced) permanent dipole $p_0 = p_0\hat{y}$.

**12.1. Force and energy of magnetic dipole in the Einstein-Laub formalism**. The force-density exerted on the moving dipole-pair is given by

$$F_{EL}(r,t) = -m_0(V/c)^2(E_0/Z_0)\delta'(x - Vt)\delta(y)\delta(z)\sin(k_0 z)\sin(\omega t)\hat{z}$$
$$+ m_0 k_0(E_0/Z_0)\delta(x - Vt)\delta(y)\delta(z)\cos(k_0 z)\sin(\omega t)\hat{x}$$
$$+ \varepsilon_0 m_0 V E_0 \delta'(x - Vt)\delta(y)\delta(z)\cos(k_0 z)\cos(\omega t)\hat{x}. \tag{80}$$

Integration over the volume of the dipole-pair now yields the net Einstein-Laub force as follows:

$$F_{EL}(t) = m_0 k_0 (E_0/Z_0)\sin(\omega t)\hat{x}. \tag{81}$$

The internal EM momentum density of the dipole pair is given by

$$\wp_{EL}^{(\text{EM})}(r,t) = E \times H_{\text{int}}/c^2 = \tfrac{1}{3}\varepsilon_0 m_0 E_0 \cos(k_0 z)\cos(\omega t)\delta(x - Vt)\delta(y)\delta(z)\hat{x}. \tag{82}$$

Differentiation with respect to time, followed by integration over the volume of the dipole-pair yields

$$F_{EL}^{(\text{int})}(t) = -\tfrac{1}{3}m_0 k_0(E_0/Z_0)\sin(\omega t)\hat{x}. \tag{83}$$

The net force experienced by the moving dipole-pair is, therefore, $\tfrac{2}{3}F_{EL}(t)$. The rate of exchange of EM energy between the field and the dipole-pair is given by

$$\frac{\partial \mathcal{E}_{EL}^{(\text{exch})}(r,t)}{\partial t} = \iiint_{-\infty}^{\infty}\left(E \cdot \frac{\partial P}{\partial t} + H \cdot \frac{\partial M}{\partial t}\right)dxdydz$$
$$= -\iiint_{-\infty}^{\infty}\varepsilon_0 m_0 E_0 V^2 \cos(k_0 z)\cos(\omega t)\delta'(x - Vt)\delta(y)\delta(z)dxdydz = 0. \tag{84}$$

The stored energy-density inside the dipole pair is given by

$$\mathcal{E}_{EL}^{(\text{int})}(r,t) = \tfrac{1}{2}\varepsilon_0 E^2 + \tfrac{1}{2}\mu_0 H^2$$
$$= \tfrac{1}{2}\varepsilon_0\left[\tfrac{2}{3}m_0 V\delta(x - Vt)\delta(y)\delta(z) - E_0\cos(k_0 z)\cos(\omega t)\right]^2$$
$$+ \tfrac{1}{2}\mu_0\left\{\left[\left(\tfrac{m_0}{3\mu_0}\right)\delta(x - Vt)\delta(y)\delta(z)\right]^2 + \left[\left(\tfrac{E_0}{Z_0}\right)\sin(k_0 z)\sin(\omega t)\right]^2\right\}. \tag{85}$$

Dropping the insignificant and constant terms from the above equation, differentiating with respect to time, then integrating over the volume of the dipole-pair, we find

$$\iiint_{-\infty}^{\infty} \frac{\partial \mathcal{E}_{EL}^{(int)}(r,t)}{\partial t} \mathrm{d}x\mathrm{d}y\mathrm{d}z = \iiint_{-\infty}^{\infty} \frac{2}{3}\varepsilon_o m_0 E_0 \omega V \delta(x-Vt)\delta(y)\delta(z)\cos(k_0 z)\sin(\omega t)\,\mathrm{d}x\mathrm{d}y\mathrm{d}z$$
$$= \tfrac{2}{3} m_0 k_0 (E_0/Z_o) V \sin(\omega t). \tag{86}$$

This is the same as the net force exerted on the dipole-pair multiplied by the dipole velocity $V$. Invoking Eq.(34), we conclude that the rest-mass $\mathfrak{m}_0$ of the dipole remains constant. As was the case in the preceding example, it is seen once again that when a dipole carries some EM momentum in its interior region, it is possible for the force and energy laws to remain consistent with each other without requiring a concomitant change in the rest mass $\mathfrak{m}_0$ of the dipole.

**12.2. Force and energy of magnetic dipole in the Lorentz formalism**. We now repeat the calculations of the preceding subsection in the Lorentz formalism, where the force-density on the dipole-pair is given by

$$\boldsymbol{F}_L(\boldsymbol{r},t) = -\varepsilon_o m_0 V E_0 \delta(x-Vt)\delta'(y)\delta(z)\cos(k_0 z)\cos(\omega t)\,\hat{\boldsymbol{y}}$$
$$+ m_0(E_0/Z_o)(1 - V^2/c^2)\delta'(x-Vt)\delta(y)\delta(z)\sin(k_0 z)\sin(\omega t)\,\hat{\boldsymbol{z}}. \tag{87}$$

Integration over the volume of the dipole-pair yields a net Lorentz force of zero on the traveling pair. Next, we evaluate the internal EM momentum in the Lorentz formalism, as follows:

$$\boldsymbol{\mathscr{p}}_L^{(EM)} = \varepsilon_o \boldsymbol{E} \times \boldsymbol{B}_{\mathrm{int}} = -\tfrac{2}{3}\varepsilon_o m_0 E_0 \cos(k_0 z)\cos(\omega t)\,\delta(x-Vt)\delta(y)\delta(z)\hat{\boldsymbol{x}}. \tag{88}$$

Differentiation with respect to time, followed by integration over the dipole volume, yields

$$\boldsymbol{F}_L^{(int)} = \iiint_{-\infty}^{\infty} \frac{\partial \boldsymbol{\mathscr{p}}_L^{(EM)}}{\partial t} \mathrm{d}x\mathrm{d}y\mathrm{d}z = \tfrac{2}{3} m_0 k_0 (E_0/Z_o)\sin(\omega t)\,\hat{\boldsymbol{x}}. \tag{89}$$

Since the Lorentz force of Eq.(87) turned out to be zero, the above internal force is the only contribution to the force exerted by the EM field on the dipole-pair. Comparison with Eqs.(81) and (83) confirms that the two methods predict the same net force on the traveling dipole-pair.

Next we calculate the energy exchange rate between the field and the dipoles in the Lorentz formalism, namely,

$$\frac{\partial \mathcal{E}_L^{(exch)}}{\partial t} = \boldsymbol{E}\cdot\left(\frac{\partial \boldsymbol{P}}{\partial t} + \mu_o^{-1}\nabla\times\boldsymbol{M}\right)$$
$$= \mu_o^{-1} m_0 E_0 \cos(k_0 z)\cos(\omega t)(1 - V^2/c^2)\delta'(x-Vt)\delta(y)\delta(z). \tag{90}$$

Integrating the above expression over the dipole volume yields a net energy exchange rate of zero. The other source of EM energy is the stored energy within the dipole pair, namely,

$$\mathcal{E}_L^{(int)}(\boldsymbol{r},t) = \tfrac{1}{2}\varepsilon_o \boldsymbol{E}\cdot\boldsymbol{E} + \tfrac{1}{2}\mu_o^{-1}\boldsymbol{B}\cdot\boldsymbol{B} =$$
$$= \tfrac{1}{2}\varepsilon_o[\tfrac{2}{3}m_0 V\delta(x-Vt)\delta(y)\delta(z) - E_0\cos(k_0 z)\cos(\omega t)]^2$$
$$+ \tfrac{1}{2}\mu_o^{-1}\{[\mu_o(E_0/Z_o)\sin(k_0 z)\sin(\omega t)]^2 + [\tfrac{2}{3}m_0\delta(x-Vt)\delta(y)\delta(z)]^2\}$$
$$= -\tfrac{2}{3}\varepsilon_o m_0 E_0 V\cos(k_0 z)\cos(\omega t)\,\delta(x-Vt)\delta(y)\delta(z) + \text{Inconsequential terms}. \tag{91}$$

Differentiation with respect to time, followed by integration over space, yields the energy exchange rate between the field and the dipole-pair as

$$\iiint_{-\infty}^{\infty} \frac{\partial \mathcal{E}_L^{(\text{int})}(r,t)}{\partial t} dxdydz = \tfrac{2}{3} m_0 k_0 (E_0/Z_o) V \sin(\omega t). \quad (92)$$

This is the same as the energy exchange rate in the Einstein-Laub formalism given by Eq.(86), confirming once again that the two formalisms predict exactly the same behavior.

**13. Slowly rotating electric dipole in an external *E*-field.** Our sixth example pertains to a point-dipole $\boldsymbol{p}_0$ sitting at the origin of the coordinates and rotating around the *y*-axis at the slow angular velocity of $\omega_0$. No relativistically-induced magnetization is produced by the rotating point-dipole. Moreover, the rotation rate is slow enough that radiation by the spinning dipole may be ignored. We express the polarization of the system as follows:

$$\boldsymbol{P}(\boldsymbol{r},t) = p_0 [\sin(\omega_0 t) \hat{\boldsymbol{x}} + \cos(\omega_0 t) \hat{\boldsymbol{z}}] \delta(x)\delta(y)\delta(z). \quad (93)$$

In the Einstein-Laub formalism, an externally applied uniform and constant electric field $\boldsymbol{E}(\boldsymbol{r},t) = E_0 \hat{\boldsymbol{x}}$ does not exert a force on the dipole, but its torque is given by

$$\boldsymbol{T}_{EL}(t) = \iiint_{-\infty}^{\infty} \boldsymbol{P} \times \boldsymbol{E} \, dxdydz = E_0 p_0 \cos(\omega_0 t) \hat{\boldsymbol{y}}. \quad (94)$$

The same torque is obtained in the Lorentz formulation, where

$$\boldsymbol{T}_L(t) = \iiint_{-\infty}^{\infty} \boldsymbol{r} \times \boldsymbol{F}_L(\boldsymbol{r},t) dxdydz = -\iiint_{-\infty}^{\infty} \boldsymbol{r} \times (\boldsymbol{\nabla}\cdot \boldsymbol{P})\boldsymbol{E} \, dxdydz = E_0 p_0 \cos(\omega_0 t) \hat{\boldsymbol{y}}. \quad (95)$$

The energy exchanged between the field and the dipole is readily calculated, as follows:

$$\frac{d\mathcal{E}^{(\text{exch})}(t)}{dt} = \iiint_{-\infty}^{\infty} \boldsymbol{E} \cdot (\partial \boldsymbol{P}/\partial t) dxdydz = E_0 p_0 \omega_0 \cos(\omega_0 t). \quad (96)$$

The *E*-field energy stored within the dipole is given by

$$\mathcal{E}^{(\text{int})}(t) = \iiint_{-\infty}^{\infty} \tfrac{1}{2}\varepsilon_o [E_0 - (p_0/3\varepsilon_o)\sin(\omega_0 t)\delta(x)\delta(y)\delta(z)]^2 dxdydz$$

$$= -\tfrac{1}{3} E_0 p_0 \sin(\omega_0 t) + \text{Inconsequential terms}. \quad (97)$$

Differentiating the above stored energy with respect to time yields $-\tfrac{1}{3} E_0 p_0 \omega_0 \cos(\omega_0 t)$, which should be added to Eq.(96) to arrive at the dipole's energy exchange rate with the external world, namely,

$$d\mathcal{E}(t)/dt = \tfrac{2}{3} E_0 p_0 \omega_0 \cos(\omega_0 t). \quad (98)$$

This rate of energy exchange between the dipole and the external world is only two thirds the product of $\boldsymbol{T}_{EL}(t)$ of Eq.(94) [or $\boldsymbol{T}_L(t)$ of Eq.(95)] and the rotation rate $\omega_0$ of the dipole. The inequality of $d\mathcal{E}(t)/dt$ and $\boldsymbol{T}(t) \cdot \boldsymbol{\omega}$ implies that the rest-mass of the dipole and, consequently, its moment of inertia, must vary with time. Unfortunately, the relativistic relations among the moment of inertia, angular velocity, (rotational) kinetic energy, angular momentum, and the rest mass are not straightforward, as they depend on the internal structure of the rotating particle. We cannot, therefore, offer a general equation similar to Eq.(34) to relate the energy imbalance $(d\mathcal{E}/dt - \boldsymbol{T} \cdot \boldsymbol{\omega})$ to the time-rate-of-change of the rest-mass (or moment of inertia) of a point-dipole. Nevertheless, it is important to recognize the necessity of accounting for the variations of the rest-mass (and moment of inertia) in order to ensure the consistency of the laws of EM force and torque with those of EM energy.

**14. Slowly rotating magnetic dipole in an external *H*-field**. In this seventh and final example, we consider a permanent point-dipole $\boldsymbol{m}_0$ sitting at the origin of the coordinates and rotating around the *y*-axis at the slow angular velocity of $\omega_0$. No relativistically-induced polarization is produced by the rotating point-dipole. Moreover, the rotation rate is slow enough that radiation by the spinning dipole may be ignored. We express the magnetization of the system as follows:

$$\boldsymbol{M}(\boldsymbol{r},t) = m_0[\sin(\omega_0 t)\,\hat{\boldsymbol{x}} + \cos(\omega_0 t)\,\hat{\boldsymbol{z}}]\delta(x)\delta(y)\delta(z). \tag{99}$$

A fundamental difference between the above magnetic dipole and the rotating electric dipole examined in the preceding section is that the magnetic dipole is always accompanied by an internal angular momentum along the axis of its magnetic dipole moment. A torque is thus needed to maintain the gyration of this angular momentum around the *y*-axis; a constant magnetic field in the *y*-direction will provide the needed (rotating) torque in the *xz*-plane. However, no energy is exchanged between the dipole and the magnetic field that is aligned with the *y*-axis. In contrast, a constant magnetic field along the *x*-axis will exchange energy with the rotating dipole while exerting a time-varying torque on the dipole along the *y*-axis. We shall assume that the dipole is properly harnessed to prevent its gyration around *x*, without interfering with its rotation around the *y*-axis.

In the Einstein-Laub formalism, an externally applied uniform and constant magnetic field $\boldsymbol{H}(\boldsymbol{r},t) = H_0\hat{\boldsymbol{x}}$ does not exert a force on the dipole, but its torque is given by

$$\boldsymbol{T}_{EL}(t) = \iiint_{-\infty}^{\infty} \boldsymbol{M} \times \boldsymbol{H}\,\mathrm{d}x\mathrm{d}y\mathrm{d}z = H_0 m_0 \cos(\omega_0 t)\,\hat{\boldsymbol{y}}. \tag{100}$$

Also in the same formalism, the energy exchanged between the field and the dipole is readily calculated, as follows:

$$\iiint_{-\infty}^{\infty} \boldsymbol{H} \cdot (\partial \boldsymbol{M}/\partial t)\mathrm{d}x\mathrm{d}y\mathrm{d}z = H_0 m_0 \omega_0 \cos(\omega_0 t). \tag{101}$$

The *H*-field energy stored within the dipole is given by

$$\mathcal{E}_{EL}^{(\mathrm{int})}(t) = \iiint_{-\infty}^{\infty} \tfrac{1}{2}\mu_0[H_0 - (m_0/3\mu_0)\sin(\omega_0 t)\,\delta(x)\delta(y)\delta(z)]^2 \mathrm{d}x\mathrm{d}y\mathrm{d}z$$

$$= -\tfrac{1}{3}H_0 m_0 \sin(\omega_0 t) + \text{Inconsequential terms}. \tag{102}$$

Differentiating the above stored energy with respect to time yields $-\tfrac{1}{3}H_0 m_0 \omega_0 \cos(\omega_0 t)$, which should be added to Eq.(101) to arrive at the dipole's energy exchange rate with the external world, namely,

$$\mathrm{d}\mathcal{E}_{EL}(t)/\mathrm{d}t = \tfrac{2}{3}H_0 m_0 \omega_0 \cos(\omega_0 t). \tag{103}$$

Now, in the Lorentz formalism, the torque exerted by the *H*-field on the spinning magnetic dipole is computed as follows:

$$\boldsymbol{T}_L(t) = \iiint_{-\infty}^{\infty} \boldsymbol{r} \times \boldsymbol{F}_L(\boldsymbol{r},t)\mathrm{d}x\mathrm{d}y\mathrm{d}z = \iiint_{-\infty}^{\infty} \boldsymbol{r} \times [(\mu_0^{-1}\boldsymbol{\nabla}\times\boldsymbol{M})\times\mu_0\boldsymbol{H}]\mathrm{d}x\mathrm{d}y\mathrm{d}z$$

$$= H_0 m_0 \cos(\omega_0 t)\,\hat{\boldsymbol{y}}. \tag{104}$$

The Lorentz torque is thus seen to be identical with the Einstein-Laub torque of Eq.(100). As for the energy exchange rate between the field and the dipole in the Lorentz formalism, we note that, since no external *E*-field acts on the dipole, the energy exchange rate $\boldsymbol{E}\cdot\mu_0^{-1}\boldsymbol{\nabla}\times\boldsymbol{M}$ vanishes. However, the stored energy inside the dipole is now given by

$$\mathcal{E}_L^{(\text{int})}(t) = \iiint_{-\infty}^{\infty} \tfrac{1}{2}\mu_0^{-1}[\mu_0 H_0 + (2m_0/3)\sin(\omega_0 t)\,\delta(x)\delta(y)\delta(z)]^2 \mathrm{d}x\mathrm{d}y\mathrm{d}z$$

$$= \tfrac{2}{3} H_0 m_0 \sin(\omega_0 t) + \text{Inconsequential terms.} \tag{105}$$

Differentiating the above stored energy with respect to time yields $\tfrac{2}{3} H_0 m_0 \omega_0 \cos(\omega_0 t)$, which is the same as that obtained in the Einstein-Laub formalism; see Eq.(103).

The two formulations thus yield the same torque and the same energy exchange rate with the external world. The energy exchange rate, however, is only two thirds the product of $\boldsymbol{T}_{EL}(t)$ of Eq.(100) [or $\boldsymbol{T}_L(t)$ of Eq.(104)] and the rotation rate $\omega_0$ of the dipole. The inequality of $\mathrm{d}\mathcal{E}(t)/\mathrm{d}t$ and $\boldsymbol{T}(t) \cdot \boldsymbol{\omega}$ implies that the rest-mass of the dipole and, consequently, its moment of inertia, must vary with time. As explained in the preceding section, we cannot offer a general formula similar to Eq.(34) to relate the energy imbalance $(\mathrm{d}\mathcal{E}/\mathrm{d}t - \boldsymbol{T} \cdot \boldsymbol{\omega})$ to the time rate of change of the rest-mass (or moment of inertia) of the point-dipole. Nevertheless, it is important to recognize the necessity of accounting for the variations of the rest-mass (and moment of inertia) in order to ensure the consistency of the laws of EM force and torque with those of EM energy.

**15. Concluding remarks**. Applying the laws of classical electrodynamics to permanent electric and magnetic dipoles moving in external EM fields, we have concluded that the rest-mass of a dipole must depend on its interaction with external fields. This is in contrast to the case of charged particles with no dipole moments, which might gain or lose kinetic energy in their interactions with external fields, but maintain a fixed rest-mass at all times.

A moving electric dipole is accompanied by a relativistically-induced magnetic dipole, and vice-versa—unless the direction of motion happens to be parallel or anti-parallel to the dipole moment, in which case the relativistically-induced companion disappears. Regardless of whether the dipole moment is inherent to the particle or is relativistically induced, the presence of a dipole moment gives rise to electric and/or magnetic fields inside the dipole, which interfere with external fields, thus producing variations in the internal energy of the particle. Equivalence of mass and energy then translates this change of the internal energy into a change of the particle's rest-mass.

In practice, variations of the rest-mass of a typical particle moving in an external EM field will be relatively small. For example, the change in the rest-mass of an electron, whose magnetic dipole moment is one Bohr magneton ($0.927 \times 10^{-23}$ J/T), when placed in a 10 Tesla external magnetic field, will be of the order of 1 mev; this is nine orders of magnitude below the electron's rest-mass of 0.511 Mev in the absence of external fields. Nevertheless, accounting for this small change in the rest-mass is absolutely essential if the Poynting theorem and the force law (whether that of Lorentz or the one due to Einstein and Laub), are to remain compatible with each other.

Finally, let us address a question that might be raised with regard to the prominent treatment of the Einstein-Laub theory in the present paper. Lack of enthusiasm for alternative theories of force and torque in classical electrodynamics is clearly understandable considering the simplicity and universality of the Lorentz force law combined with its tremendous success in explaining the observed EM phenomena. In fact, the Lorentz formulation (in conjunction with Maxwell's equations) is how macroscopic electrodynamics is almost universally understood today.

There exist, however, at least two good reasons why the Einstein-Laub theory deserves a re-examination. First, from a purely aesthetic point of view, the Lorentz formulation in the presence of magnetic matter requires the notions of "hidden energy" and "hidden momentum" in order to comply with the conservation laws and with special relativity. In contrast, the Einstein-Laub method eschews hidden entities under similar circumstances. The initial simplicity of the Lorentz formalism—a major source of its appeal—is thus marred by subsequent complications arising from the need to keep track of hidden energy and hidden momentum.

The second incentive for re-examining the Einstein-Laub theory is that its predictions of force-density and torque-density *distributions* inside matter are at variance with those of the Lorentz formalism. We have described these differences in some detail in a recent paper [23]. Briefly, while the *total* force (and also *total* torque) exerted by EM fields on isolated objects are *always* the same in the two formulations, their predicted force and torque *distributions* inside material media exhibit substantial differences. Consequently, the possibility exists that these alternative theories could be distinguished from each other in radiation pressure experiments involving deformable media. As pointed out in [23], the existing experimental evidence (e.g., Ashkin and Dziedzic's focused light experiments on the surface of water [66], and also electrostriction effects in nonlinear optics [67]) is supportive of the Einstein-Laub theory.

We emphasize that the Einstein-Laub formulation provides a complete and consistent theory of electrodynamics. It incorporates Maxwell's macroscopic equations (*without* the need to introduce local averaging of electric and magnetic dipoles), is consistent with the conservation laws of energy, linear momentum, and angular momentum, and also complies with special relativity. Its fundamental fields are *not* $E$ and $H$, but rather the ($E$, $B$) pair and the ($D$, $H$) pair. The Einstein-Laub formalism is *not* limited to "the archaic treatment of magnetism" or "magnetostatics," as some have suggested; rather, it applies to every problem in classical electrodynamics. Serious students of physics will have no reason to worry about the possibility of confusion if the Einstein-Laub formulation turns out to be the correct (i.e., experimentally verified) theory of electrodynamics. The above statements are *not* merely personal opinions; we have provided rigorous proofs of them throughout the present paper and in previous publications.

## Appendix A

To derive Eq.(34), we start from the relativistic formulas for the energy-density $\mathcal{E}(\mathbf{r},t)$ and momentum-density $\boldsymbol{p}(\mathbf{r},t)$ of a point-particle of rest-mass $\mathfrak{m}_o(t)$ and velocity $\boldsymbol{V}_p(t)$ given in Eqs.(26-28). Considering that $\gamma'(t) = \gamma^3 \boldsymbol{V}_p \cdot \boldsymbol{V}'_p / c^2$, we find

$$\frac{\mathrm{d}\mathcal{E}(t)}{\mathrm{d}t} = \iiint_{-\infty}^{\infty} \frac{\partial \mathcal{E}(\mathbf{r},t)}{\partial t} \mathrm{d}x \mathrm{d}y \mathrm{d}z = \mathfrak{m}'_o(t)\gamma(t)c^2 + \mathfrak{m}_o(t)\gamma'(t)c^2$$

$$= \gamma \left[ \frac{\mathrm{d}(\mathfrak{m}_o c^2)}{\mathrm{d}t} + \mathfrak{m}_o \gamma^2 \boldsymbol{V}_p \cdot \boldsymbol{V}'_p \right]. \tag{A1}$$

Similarly,

$$\boldsymbol{F}_{\mathrm{ext}}(t) = \iiint_{-\infty}^{\infty} \frac{\partial \boldsymbol{p}(\mathbf{r},t)}{\partial t} \mathrm{d}x \mathrm{d}y \mathrm{d}z$$

$$= \mathfrak{m}'_o(t)\gamma(t)\boldsymbol{V}_p(t) + \mathfrak{m}_o(t)\gamma'(t)\boldsymbol{V}_p(t) + \mathfrak{m}_o(t)\gamma(t)\boldsymbol{V}'_p(t)$$

$$= \gamma \left[ \frac{\mathrm{d}(\mathfrak{m}_o c^2)}{\mathrm{d}t}(\boldsymbol{V}_p/c^2) + \mathfrak{m}_o \gamma^2 (\boldsymbol{V}_p \cdot \boldsymbol{V}'_p)(\boldsymbol{V}_p/c^2) + \mathfrak{m}_o \boldsymbol{V}'_p \right]. \tag{A2}$$

Multiplying the above equation into $\boldsymbol{V}_p(t)$ yields

$$\boldsymbol{F}_{\mathrm{ext}}(t) \cdot \boldsymbol{V}_p(t) = \gamma \left\{ \frac{\mathrm{d}(\mathfrak{m}_o c^2)}{\mathrm{d}t}(\boldsymbol{V}_p \cdot \boldsymbol{V}_p/c^2) + \mathfrak{m}_o (\boldsymbol{V}_p \cdot \boldsymbol{V}'_p)[\gamma^2 (\boldsymbol{V}_p \cdot \boldsymbol{V}_p/c^2) + 1] \right\}$$

$$= \gamma \left[ \frac{\mathrm{d}(\mathfrak{m}_o c^2)}{\mathrm{d}t}(\boldsymbol{V}_p \cdot \boldsymbol{V}_p/c^2) + \mathfrak{m}_o \gamma^2 \boldsymbol{V}_p \cdot \boldsymbol{V}'_p \right]. \tag{A3}$$

We now subtract Eq. (A3) from Eq. (A1) to arrive at

$$\gamma \left[\frac{d\mathcal{E}(t)}{dt} - \boldsymbol{F}_{\text{ext}}(t) \cdot \boldsymbol{V}_p(t)\right] = \gamma^2(1 - \boldsymbol{V}_p \cdot \boldsymbol{V}_p/c^2)\frac{d(\mathfrak{m}_o c^2)}{dt} = \frac{d(\mathfrak{m}_o c^2)}{dt}, \tag{A4}$$

which is the same as Eq. (34).

## Appendix B

Consider a small particle having polarization $\boldsymbol{P}(\boldsymbol{r},t)$ and magnetization $\boldsymbol{M}(\boldsymbol{r},t)$, traveling in free space. The fields inside the particle are $\boldsymbol{E}_{\text{in}}(\boldsymbol{r},t)$ and $\boldsymbol{H}_{\text{in}}(\boldsymbol{r},t)$, while the fields outside are $\boldsymbol{E}_{\text{out}}(\boldsymbol{r},t)$ and $\boldsymbol{H}_{\text{out}}(\boldsymbol{r},t)$. Defining the Poynting vector as $\boldsymbol{S}(\boldsymbol{r},t) = \boldsymbol{E}(\boldsymbol{r},t) \times \boldsymbol{H}(\boldsymbol{r},t)$, one may write the Poynting theorem as follows [see Eq. (9)]:

$$\boldsymbol{\nabla} \cdot \boldsymbol{S}(\boldsymbol{r},t) + \frac{\partial}{\partial t}(\tfrac{1}{2}\varepsilon_0 \boldsymbol{E} \cdot \boldsymbol{E} + \tfrac{1}{2}\mu_0 \boldsymbol{H} \cdot \boldsymbol{H}) + \left(\boldsymbol{E} \cdot \frac{\partial \boldsymbol{P}}{\partial t} + \boldsymbol{H} \cdot \frac{\partial \boldsymbol{M}}{\partial t}\right) = 0. \tag{B1}$$

Imagine a closed surface immediately outside and surrounding the particle at time *t*. Gauss's theorem applied to Eq. (B1) yields

$$\oiint \boldsymbol{S}(\boldsymbol{r},t) \cdot d\boldsymbol{\sigma} + \frac{d}{dt}\iiint(\tfrac{1}{2}\varepsilon_0 \boldsymbol{E}_{\text{in}} \cdot \boldsymbol{E}_{\text{in}} + \tfrac{1}{2}\mu_0 \boldsymbol{H}_{\text{in}} \cdot \boldsymbol{H}_{\text{in}})\, dv + \iiint\left(\boldsymbol{E}_{\text{in}} \cdot \frac{\partial \boldsymbol{P}}{\partial t} + \boldsymbol{H}_{\text{in}} \cdot \frac{\partial \boldsymbol{M}}{\partial t}\right)dv = 0. \tag{B2}$$

The first term on the left-hand side of Eq. (B2) is the surface integral over the closed surface surrounding the particle. The remaining integrals are taken over the volume of the particle. A similar equation can be written for the region outside the particle. Taking into account the vanishing of the integral of $\boldsymbol{S}(\boldsymbol{r},t)$ over a closed surface at infinity, we write

$$-\oiint \boldsymbol{S}(\boldsymbol{r},t) \cdot d\boldsymbol{\sigma} + \frac{d}{dt}\iiint\left(\tfrac{1}{2}\varepsilon_0 \boldsymbol{E}_{\text{out}} \cdot \boldsymbol{E}_{\text{out}} + \tfrac{1}{2}\mu_0 \boldsymbol{H}_{\text{out}} \cdot \boldsymbol{H}_{\text{out}}\right) dv = 0. \tag{B3}$$

Adding Eq. (B3) to Eq. (B2) now yields

$$\iiint_{\text{particle}} \left(\boldsymbol{E}_{\text{in}} \cdot \frac{\partial \boldsymbol{P}}{\partial t} + \boldsymbol{H}_{\text{in}} \cdot \frac{\partial \boldsymbol{M}}{\partial t}\right) dv = -\frac{d}{dt}\iiint_{\text{all space}}(\tfrac{1}{2}\varepsilon_0 \boldsymbol{E} \cdot \boldsymbol{E} + \tfrac{1}{2}\mu_0 \boldsymbol{H} \cdot \boldsymbol{H})\, dv. \tag{B4}$$

In the example of a point electric dipole moving away from a charged wire in Sec. 9, we have $\boldsymbol{M} = 0$ and $\boldsymbol{H}_{\text{in}} = 0$. Therefore, Eq. (39) evaluates the left-hand-side of Eq. (B4), which, as the right-hand-side of Eq. (B4) indicates, is equal in magnitude and opposite in sign to the time-rate-of-change of the *total* energy of the EM field. It is thus seen that the calculation of *total* field energy stored inside as well as outside the dipole is not necessary; the use of the Poynting theorem has eliminated the need to evaluate the integral on the right-hand-side of Eq. (B4). In other words, Eq. (39) of Sec. 9, which evaluates the left-hand side of Eq. (B4), yields the time-rate-of-change of the *total* EM field energy. Since the force exerted by the external *E* field on the point-dipole is given by Eq. (41), it is tempting to conclude that Eqs. (39) and (41) satisfy the energy conservation relation $\boldsymbol{F}_{\text{ext}} \cdot \boldsymbol{V}_p = -d\mathcal{E}_{\text{total}}/dt$.

The story, however, does not end here. It is important to recognize the dipole in Sec. 9 as a small solid sphere with an internal *E* field—as opposed to a pair of positive and negative charges attached to the opposite ends of a tiny stick. Therefore, one must add to Eq. (39) the time-rate-of-change of the energy stored *inside* the dipole, namely, $\mathcal{E}(t)$ of Eq. (38). Subsequently, this latter contribution to the rate-of-change of energy, which is *not* cancelled out by $\boldsymbol{F}_{\text{ext}} \cdot \boldsymbol{V}_p$, produces the time-rate-of-change of the rest-mass $\mathfrak{m}_o$ given by Eq. (42). Clearly, one must be careful to account for *all* contributions to the energy of the particle, not just the obvious ones.

In the system depicted in Fig. 2, as the dipole recedes from the charged wire, its internal energy $\mathcal{E}(t)$ given by Eq. (38) varies with time (due to the declining magnitude of the external field). Recalling the equivalence of mass and energy, a change in the internal energy of the spherical point-particle must result in a change of its rest-mass. Indeed, Eq. (42) is nothing but the time-derivative of the internal energy given by Eq. (38) — with the factor $\gamma$ disappearing from Eq. (42) due to the time-dilation in going from the rest frame to the laboratory frame.